# Label-free pathological subtyping of non-small cell lung cancer using deep classification and virtual immunohistochemical staining

Zhenya Zang[1], David A Dorward[2], Katherine E Quiohilag[2], Andrew DJ Wood[2], James R Hopgood[3], Ahsan R Akram[1†*], Qiang Wang[1†*]

[1]Centre for Inflammation Research, Institute of Regeneration and Repair, The University of Edinburgh, Edinburgh, UK.

[2]Department of Pathology, Royal Infirmary of Edinburgh, Edinburgh, UK.

[3]Institute of Imaging, Data and Communications, School of Engineering, The University of Edinburgh, Edinburgh, UK.

† These authors jointly supervised the work and are co-senior authors

* Corresponding authors: Q.Wang@ed.ac.uk, Ahsan.Akram@ed.ac.uk

## Abstract

The differentiation between pathological subtypes of non-small cell lung cancer (NSCLC) is an essential step in guiding treatment options and prognosis. However, current clinical practice relies on multi-step staining and labelling processes that are time-intensive and costly, requiring highly specialised expertise. In this study, we propose a label-free methodology that facilitates autofluorescence imaging of unstained NSCLC samples and deep learning (DL) techniques to distinguish between non-cancerous tissue, adenocarcinoma (AC), squamous cell carcinoma (SqCC), and other subtypes (OS). We conducted DL-based classification and generated virtual immunohistochemical (IHC) stains, including thyroid transcription factor-1 (TTF-1) for AC and p40 for SqCC. We evaluated these methods using two types of autofluorescence imaging: intensity imaging and lifetime imaging. The results demonstrate the exceptional ability of this approach for NSCLC subtype differentiation, achieving an area under the curve above 0.981 and 0.996 for binary- and multi-class classification. Furthermore, this approach produces clinical-grade virtual IHC staining, which was blind-evaluated by three experienced thoracic pathologists. Our label-free NSCLC subtyping approach enables rapid and accurate diagnosis without the need for conventional tissue processing and staining. Both strategies can significantly accelerate diagnostic workflows and support efficient lung cancer diagnosis, without compromising clinical decision-making.

## Introduction

Lung cancer remains the most commonly occurring and leading cause of cancer-associated mortality globally, accounting for 12.4% of all cancer diagnoses and 18.7% of cancer-related deaths[1]. Pathological diagnosis, subtyping, and molecular phenotyping are central to effectively managing the disease and informing prognosis. Non-small cell lung cancer (NSCLC) accounts for approximately 80% of newly diagnosed lung cancers[2,3], with adenocarcinoma (AC) and squamous cell carcinomas (SqCC) comprising approximately 50% and 30% of NSCLC cases, respectively[4]. However, distinguishing between these and other subtypes based on morphological features alone can be challenging due to the loss of distinct histological differences in more poorly differentiated carcinomas. Immunohistochemical (IHC) staining is therefore frequently employed to aid in phenotypic classification, but this process requires additional time, labour, and cost, which can impact timely diagnosis and delay treatment decisions. Furthermore, the additional tissue sections required for IHC risk exhausting the limited available cellular material, which is also necessary for downstream DNA and RNA-based molecular profiling, thereby increasing the risk of requiring a repeat biopsy. Given advancements in computational power, data-driven algorithms, and efficient imaging modalities that explore cellular function and morphology, a fast and accurate computer-aided classification solution is highly desirable to enhance diagnostic efficiency and facilitate the rapid diagnosis of label-free, stain-free tissues.

Deep neural networks (DNNs) have emerged as powerful tools for pattern recognition and have been widely applied to both haematoxylin and eosin (H&E) and IHC sections to aid in lung cancer classification over the past decade. Coudray *et al.*[5] leveraged deep learning (DL) to classify three subtypes, AC, SqCC, and normal tissue, as well as six AC mutation classes using datasets from The Cancer Genome Atlas (TCGA), including H&E-stained images of lung cancer. Similarly, Noorbakhsh *et al.*[6] utilised whole slide images (WSIs) of H&E-stained tissue from the same TCGA dataset to classify AC and SqCC, employing an Inception v3 architecture. Chen *et al.*[7] proposed a classification strategy for AC, SqCC, and non-cancerous tissue using a ResNet-50 architecture trained on H&E WSIs from multiple pathology departments and tested on TCGA datasets. Multiple DNN architectures[8] have also been employed to identify AC and SqCC transcriptomic subtypes and distinguish tumour regions from adjacent benign tissue. Sadhwani *et al.*[9] developed a convolutional neural network (CNN) trained on histological features to classify histologic patterns in AC from WSIs of H&E-stained tissue and predict tumour mutation burden. Kanavati *et al.*[10] combined a CNN with a recurrent neural network (RNN) to classify AC, SqCC, small-cell lung cancer (SCLC), and non-neoplastic tissues using patched images from WSIs of H&E-stained tissues. Diff-Quik-stained lung WSIs were used[11] to train an attention-based DL model for performing holistic six-class discrimination, including benign, AC, SqCC, NSCLC-not otherwise specified, small cell lung cancer, and other malignancies. However, to date, most DL-based subtyping methods are based on assessing some form of stained tissue.

Label-free autofluorescence imaging utilises the intrinsic fluorescence of biological tissues for cancer diagnosis, capturing metabolic and structural changes at the cellular level. One of the widely used endogenous signals is autofluorescence intensity, which has been utilised for the detection and diagnosis of various cancers, including lung[12], oral[13], breast[14], prostate[15], and skin[16] cancer. Another critical feature of autofluorescence signals is lifetime, characterised by a fluorophore's decay from the excited state to the ground state[17]. Fluorescence lifetime imaging microscopy (FLIM) can capture this unique feature to investigate subtle changes in the biological environment at a molecular level[17]. Fluorescence lifetime has broad applications in cancer diagnosis, including lung cancer[18, 19], prostate cancer[20], breast cancer[21], and skin cancer[22]. Due to its capability at the molecular level, fluorescence lifetime can also be used to differentiate cell types and phenotypes, such as T-cell activation[23], cancer cell phenotypes[24], and macrophage subtypes[25]. Despite the use of label-free signals for lung cancer detection, the effectiveness of these features for lung cancer subtyping, or cancer subtyping in general, remains uncertain, particularly in addressing interpatient heterogeneity. Our recent research has demonstrated the feasibility of combining DL techniques with FLIM images for lung cancer diagnosis[26, 27, 28]. Furthermore, we have managed to translate FLIM images into virtual H&E images across multiple tumour types[29]. Both allow for timely and accurate lung cancer detection without requiring the conventional tissue processing and staining procedures. All these indicate the potential of label-free signals for advanced cancer characterisation, with the integration of DL for improved fidelity and reduced tissue consumption. Recent advances in virtual histological staining provide promising alternatives to conventional cancer pathology, allowing rapid digital staining with clinical-grade quality[34, 35]. Generally, virtual staining techniques can be categorised into two groups: label-free virtual staining and stain-to-stain (S2S) transformation. In the label-free domain, autofluorescence images[36,37,38,39], bright-field images[34,40], FLIM images[41,29], and photoacoustic images[42,43] have been used as inputs to synthesise H&E and IHC images (HER2[39], SOX10[44], FAP-CK[45], etc) for different organ types. For S2S approaches, H&E-stained tissue is mainly used as input to synthesise other types of stains, with proteins of interest marked by specific biomarkers such as Fibroblast Activation Protein and Cytokeratin[46], oestrogen receptor and Anti-Prosurfactant Protein[47], Periodic Schiff-Methenamine[48], etc. In addition to one-to-one S2S methods, a multiplexed virtual stain approach[49] can translate H&E images into high-fidelity IHC images of different markers. More details of the two types of virtual stain methods are summarised in a review article[34]. To date, no virtual staining approach has targeted the proteins used to characterise AC and SqCC in routine clinical practice, namely, Thyroid Transcription Factor 1 (TTF-1)[50] and p40, respectively[51, 52].

Considering this, we propose two strategies: the NSCLC classifier and virtual staining, to distinguish between major NSCLC subtypes, and validate both using label-free intensity and lifetime images. For the NSCLC classifier, we apply various DL models to predict normal tissue, AC, SqCC, and other NSCLC subtypes, utilising widely adopted metrics to assess performance. For virtual staining, we use a generative adversarial network (GAN), previously employed for virtual H&E staining[29], to generate synthetic TTF-1- and p40 images for AC and SqCC, respectively. To assess the quality of the generated virtual IHC images, we conducted a blind evaluation by three certified pathologists, complemented by quantitative analysis.

# Results

## Performance in Binary NSCLC Subtype Classification

We evaluated our approach using 631 tissue microarray (TMA) cores from more than 280 patients. This included non-cancerous lung, AC, SqCC, and other subtypes (OS), and encompassed a variety of pathological stages (Clinical details in Supplementary Table 1). A unique feature of this dataset is the ability to perform confirmatory IHC staining for markers of interest on the same TMA core (due to label-free imaging), enabling perfect co-registration. Intensity images were contrast-enhanced and fed into DL models in single-channel greyscale. In contrast, lifetime images were processed to generate four-channel RGB images, as this format proved optimal for lifetime-based classification[28]. A Python script generates 224 × 224 patches from entire large core images (approximately 5,000 × 5,000 pixels) for the DL model. Supplementary Table 2 summarises the number of cores and patches for each subtype across training, validation, and test sets. Since the number of OS and normal tissue cores was smaller than that of AC and SqCC, a 30% horizontal and vertical overlap was applied when patching the normal and OS cores to balance the dataset and stabilise the training process.

We first evaluated the performance on binary classification across four label groups: cancer vs. non-cancerous, AC vs. (SqCC + OS), SqCC vs. OS, and AC vs. SqCC. A quantitative evaluation of DL models trained with FLIM and intensity images is shown in Fig. 1 and Supplementary Fig. 1. Fig. 1a illustrates the binary classification workflow, in which samples are first classified into cancerous and non-cancerous classes. The cancerous samples are then further classified into specific subtypes. For intensity-based classification, shown in Fig. 1b, the binary classification of cancer vs. non-cancer achieves nearly perfect ROC curves and AUC scores. In contrast, the classification of AC vs. (SqCC + OS) and SqCC vs. OS shown in Fig. 1d and 1f, yields slightly lower AUC values. The corresponding confusion matrices align with the ROC curve results.

Approximately 12.60% and 13.36% of patches were misclassified for AC vs. (SqCC + OS), and 5.41% and 19.51% for SqCC vs. OS, compared to only 0.30% for both cancer vs. non-cancer classifications. The FLIM image-trained DL model, shown in Fig. 1c, 1e, and 1g also generate accurate subtyping results, with perfect classification of cancer vs. non-cancer, and near-perfect classification of SqCC versus OS. Approximately 6.27% and 7.54% of patches were misclassified in AC vs. (SqCC + OS), while 3.97% and 0.77% were misclassified in SqCC vs. OS. In contrast, cancer vs. non-cancer classification achieved significantly lower error rates of 0.25% and 0.09%. Distinguishing AC from SqCC, as shown in Fig. 1h and i, which has higher clinical relevance, achieves relatively lower AUCs than our other classification tasks; yet it remains high relative to state-of-the-art methods. Notably, the enhanced performance from FLIM could be attributed to the microenvironmental information provided by FLIM images, where metabolic differences and proteomic functions contribute to more discriminative feature representations. For example, SqCC tumours often exhibit higher glucose uptake and reliance on oxidative phosphorylation, whereas AC tumours favour aerobic glycolysis[53]. This metabolic reprogramming leads to differences in the fluorescence lifetime of endogenous fluorophores, such as nicotinamide adenine dinucleotide plus hydrogen (NADH) and flavin adenine dinucleotide (FAD). Proteomic analyses also revealed that AC and SqCC tumours differ in mitochondrial content and function[54]. SqCC tumours often have higher mitochondrial activity, correlating with increased oxidative metabolism. These differences can influence the fluorescence lifetime of mitochondria-associated fluorophores.

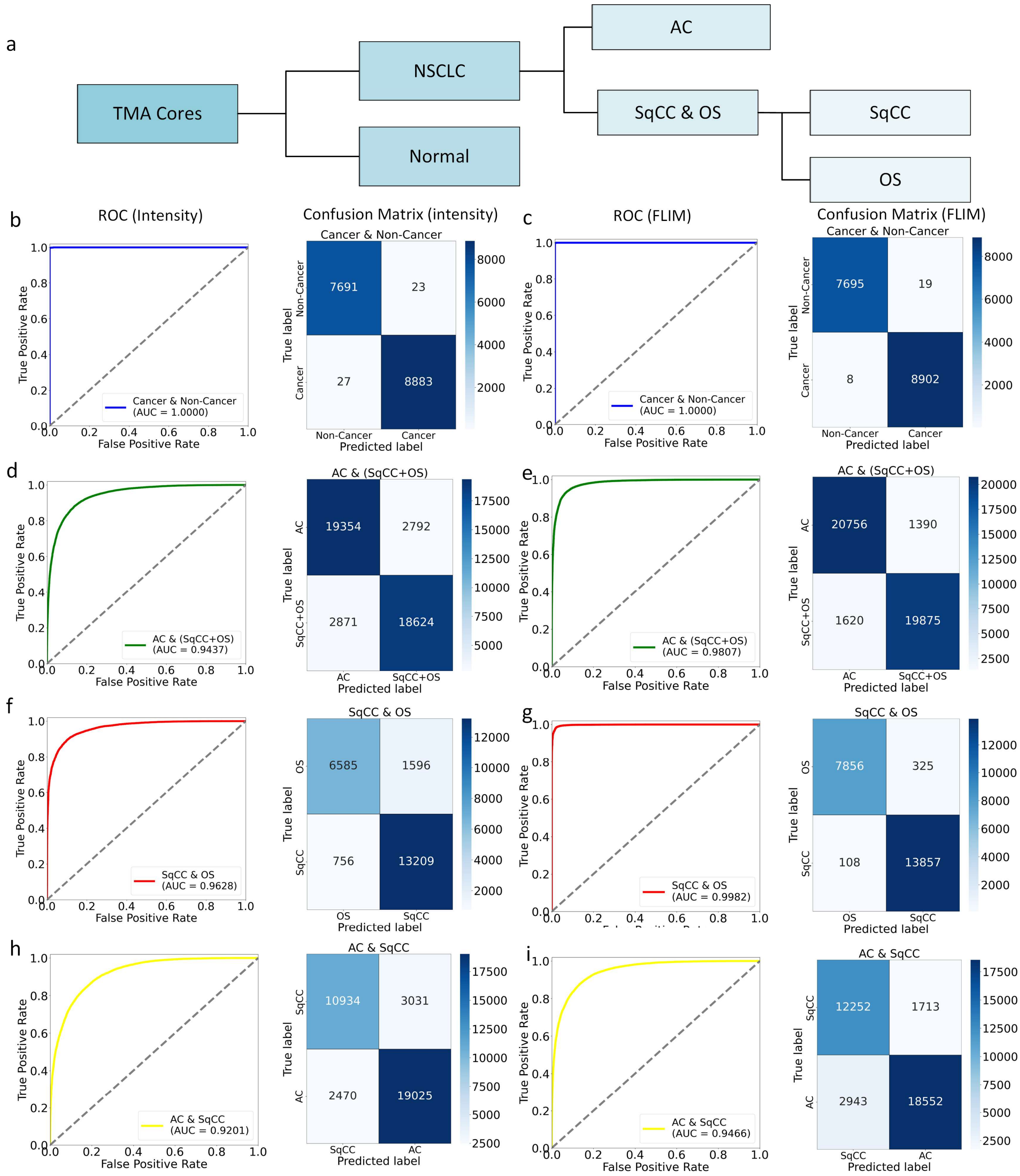

***Figure 1. Binary classification performance evaluation for three groups of cancer types: Cancer vs. Non-Cancer, AC vs. (SqCC + OS), and SqCC vs. OS.*** *(a) Subtyping overview. (b), (d), and (f) show ROC curves with AUC scores and confusion matrices for the three binary classifications based on intensity images. (c), (e) and (g) present the same evaluation metrics based on FLIM images, (h) and (i) present ROCs and confusion matrices subtyping AC and SqCC from intensity- and FLIM-based models.*

Besides statistical evaluations of patch images, we visualise representative cores of AC, SqCC, and OS, along with their corresponding intensity and FLIM images, in Fig. 2. Additionally, H&E-stained images, a standard modality used by pathologists for screening, are also provided. To visualise the prediction probabilities from intensity- and FLIM-based DL models for each patch within a core, probabilistic maps are shown for all three subtypes. Each square in the probability maps represents a 224 × 224 image processed by the intensity- and FLIM-based DL models. Both models correctly classify most patches; however, the intensity-based model distinguishes malignant from benign regions, with the latter appearing as white patches. In contrast, the FLIM-based model produces more uniform probability maps, classifying nearly all patches as cancerous, irrespective of tissue composition. This suggests that non-malignant tissue components (e.g., stroma and inflammatory cells) may be influenced by adjacent tumours, affecting their lifetime characteristics. Supplementary Fig. 2–4 present five representative patches with varying tissue components, morphologies, and corresponding predicted probabilities. Notably, patches with fewer cancer cells (e.g., index 2 and 5 in Supplementary Fig. 2; index 2 in Supplementary Fig. 3; and index 1, 2, and 5 in Supplementary Fig. 4) receive lower probabilities from the intensity-based model, indicating a non-malignant classification. However, as the entire cores were pathologically labelled as malignant, the FLIM-based model assigns higher probabilities, suggesting that, beyond cancer cell morphology in intensity and H&E-stained images, additional tissue components in FLIM images may serve as valuable indicators for cancer detection when combined with DL-based approaches. To quantify the discrepancy between the two models, Supplementary Fig. 2–4 show the distributions of predicted probabilities for patches from three representative cores. The FLIM-based model exhibits a sharper distribution concentrated near higher probabilities, suggesting more confident and accurate predictions than the intensity-based model. Besides the resultant probabilities shown in Supplementary Fig. 2–4, the interpretability of the DL model and visualisation are demonstrated using Grad-cam++[55], which is widely used in DL-based medical imaging. Grad-CAM++ highlights the spatial regions most influential for each prediction, revealing modality-specific differences in feature usage. Rows f and h in Supplementary Fig. 2–4 show the activation heatmaps from the intensity- and FLIM-based models. In adenocarcinoma and SqCC (Supplementary Figs. 2 and 3), FLIM-derived heatmaps exhibit stronger, more coherent signals, indicating richer feature extraction compared with intensity alone. The FLIM model consistently localises tumour regions and avoids background or stromal areas (e.g., index-3 in Supplementary Fig. 2f, h and index-2 in Supplementary Fig. 3f, h), whereas the intensity model occasionally focuses on background (index-2 in Supplementary Fig. 3f, h) or stroma (index-3 in Supplementary Fig. 4f, h). Salience maps facilitate the identification of cancerous regions and provide a transparent way to validate that the model's decisions align with recognised diagnostic criteria, supporting safe translation into clinical workflows.

To further investigate how the models process FLIM and intensity images, we visualised the distribution of feature maps. Fig. 2d–e present t-SNE[55] cluster plots of the feature embeddings from the last fully connected layer, illustrating the clustering of inferred patches across all test cores, from intensity- and FLIM-based models. Both models exhibit well-defined clusters, particularly for patches from normal cores, while the FLIM-based model shows slightly more distinct separation for the OS class. However, sparse outliers between clusters, particularly for AC, normal, and OS in Fig. 2d, highlight areas where classification challenges persist. Individually colour-coded t-SNE clusters for the four classes are also presented to intuitively visualise the inter-class discrimination capabilities in the feature space of our intensity- and FLIM-based models, as shown in Fig. 2e and Fig. 2g. Consistent with the binarised t-SNE plots, both intensity- and FLIM-based models exhibit clear classification boundaries, with the FLIM-based model showing sharper separation. ACC and SqCC classes show greater overlap than the other subtypes in both models.

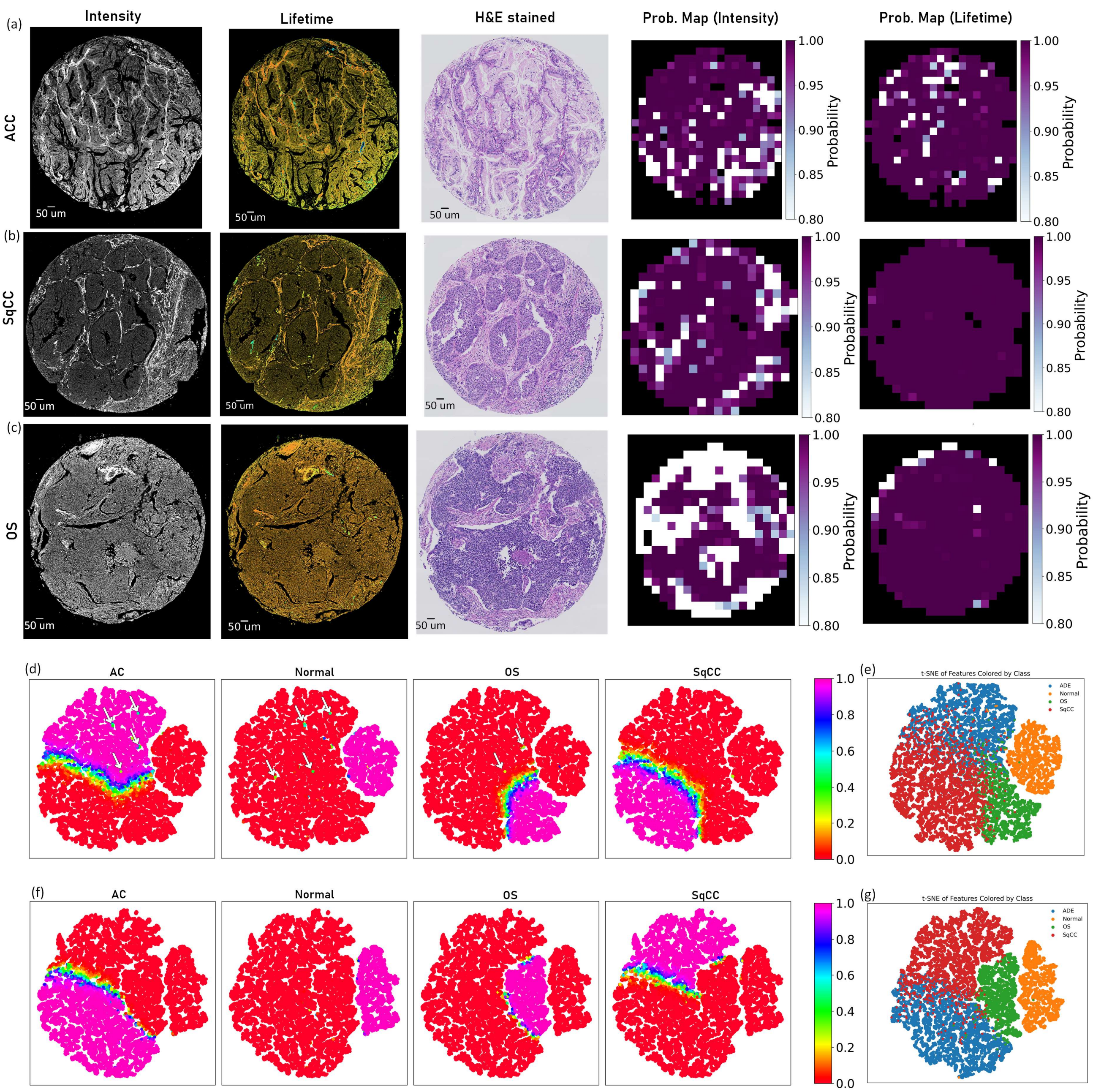

***Figure 2.*** *(a–c) Representative AC, SqCC, and OS cores are shown through intensity, FLIM, and H&E-stained images, along with probability maps inferred from intensity-based and FLIM-based models. The colour bar range spans from 0.8 to 1.0, with patches having probabilities below 0.8 displayed in white. (d–g) patch-based t-SNE clustering results derived from the last fully connected layer of the DNN models trained with (d) intensity and (f) FLIM images are presented. Both the intensity- and FLIM-based models exhibit four distinct class clusters with clear boundaries. Misclassified outliers from*

*the intensity-based model are indicated by white arrows in (d), corresponding to (e) and (g), which show the colour-coded t-SNE plots of each class cluster for the intensity- and FLIM-based models, respectively.*

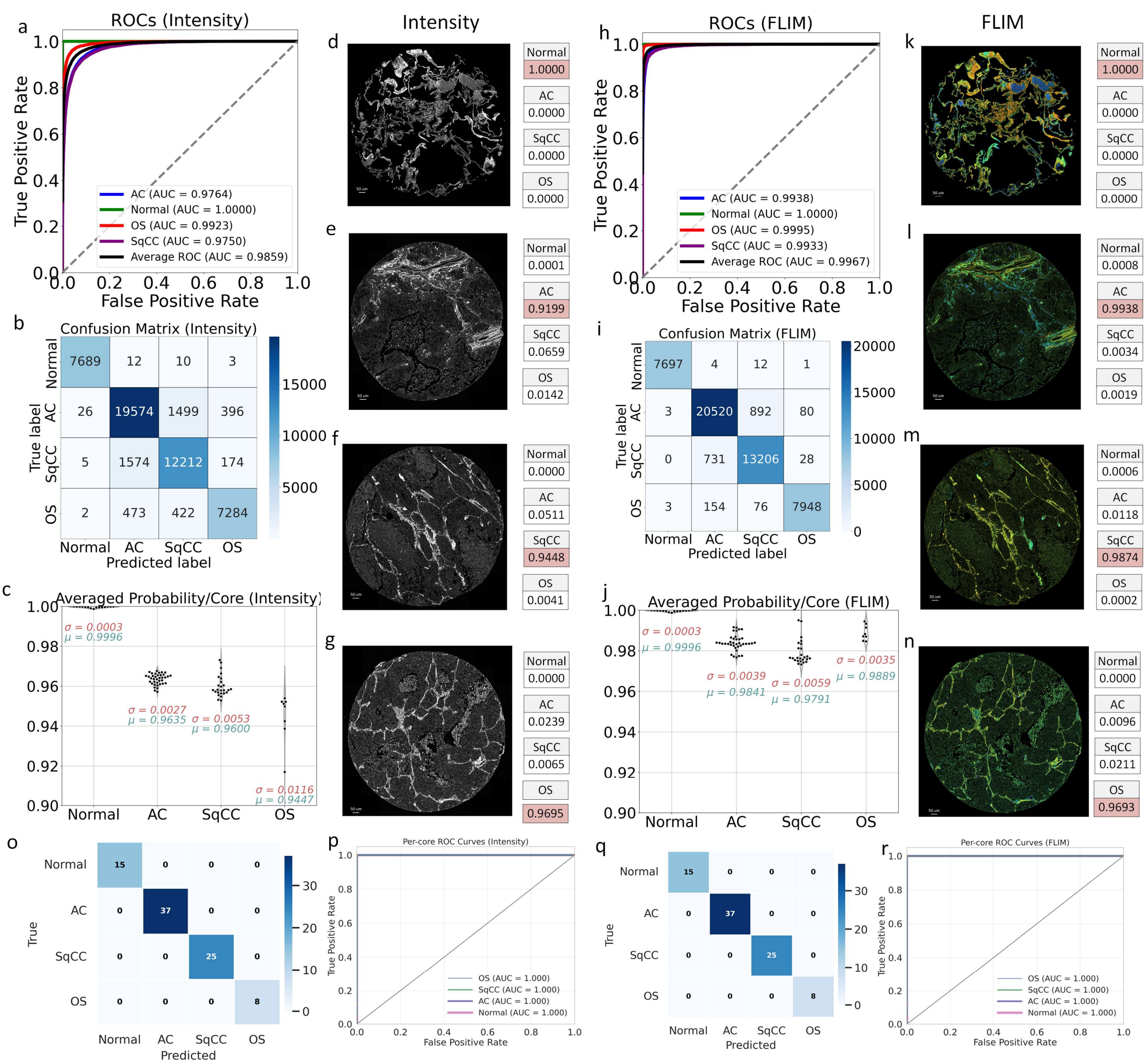

***Figure 3. Evaluation of the DL model for four-class multi-classification using FLIMs and intensity image datasets.*** *(a)-(c) and (h)-(j) indicate ROCs, confusion matrices, and core-based prediction probabilities (indicated by violin plots with means μ and standard deviations σ) from the DL model trained with intensity and FLIM images, respectively. (d)-(g) and (k)-(n) are examples of intensity and FLIM images of cases normal, AC, SqCC, and OS, indicated with corresponding four-class prediction probabilities. (o) and (p), and (q) and (r) are the core-based confusion matrices and ROC curves from the intensity- and FLIM-based models, respectively.*

## Performance in Multi-Class NSCLC Subtype Classification

Besides binary classification, we also trained the DL models for multi-class classification with four label categories using the same dataset. For patch-based evaluation, Fig. 3a and 3h present the ROC curves for each class, demonstrating strong performance across all categories, with the highest ROC curves and AUC scores for normal and OS. The FLIM-based model (Fig. 3h) achieves consistently high AUC scores (greater than 0.99) across all classes, while the confusion matrix (Fig. 3i) suggests a slightly lower misclassification rate compared to the intensity-based model (Fig. 3b). Fig. 3d–g and 2k–n showcase the representative intensity and FLIM images for normal, AC, SqCC, and OS cases, along with the predicted probabilities. While both models achieve strong classification performance, distinguishing between AC and SqCC remains the most challenging. Core-based evaluation is presented in Fig. 3o–r, where the classification probabilities for each core are averaged across all corresponding patches, resulting in correct predictions for all testing cores.

Alongside patch-based classification, we also statistically evaluated the accuracy of the FLIM- and intensity-based models for each class core. A similar patching strategy to the patch-based classification was applied to the core-based classification. After inferring each patch, the probabilities were appended and averaged to produce a final probability indicating the core's class. Fig. 3c and 3j demonstrate the distributions of probabilities, mean values ($\mu$), and standard deviations ($\sigma$) of cores for the four subtypes. Both models accurately classify OS, achieving $\mu$ values close to 1.00 and small $\sigma$ values. The FLIM-based model outperforms the intensity-based model for the other three subtypes, with higher $\mu$ and lower $\sigma$ values. These results further demonstrate that although the topological features provided by intensity images can ensure high classification accuracy and precision, functional features from FLIM images are also crucial for further enhancing performance. We also trained and tested ResNet[56] and EfficientNet[57] to compare with DenseNet[58]. The results show that DenseNet performs best across various evaluation metrics. The detailed comparisons are documented in Supplementary Table 3.

Our label-free subtyping provides timely classification references. The classification of normal tissue, AC, SqCC, and OS types takes 1.79, 2.64, 2.32, and 2.64 seconds per core, respectively, on an NVIDIA RTX A5000. Each patch requires approximately 8.5 ms for classification. Normal tissues consume less time as the cores are sparse, and more patches are filtered out using a pre-defined signal-to-background ratio (SBR). The inference time per patch is 0.012 seconds with the intensity-based model and 0.013 seconds with the FLIM-based model. For multi-class classification, the inference time per patch is 0.015 seconds with the intensity-based model and 0.016 seconds with the FLIM-based model.

## Virtual IHC staining for NSCLC subtyping

We evaluated virtual TTF-1 staining on an independent cohort containing 9 TMA cores, including 3 lung AC, 4 lung SqCC, and 2 other NSCLC subtypes. We also conducted a similar evaluation of virtual p40 staining using a separate cohort of 10 TMA cores for testing, which included 4 AC, 4 SqCC, and 2 OS. To comprehensively evaluate the quality of virtual staining and its clinical suitability, eight cases from each virtual staining method were scored by three experienced thoracic pathologists. The evaluation focused on overall staining quality and its utility in diagnosing AC and SqCC. For staining details, tumour cells were assessed for the presence and accuracy of staining, using corresponding H&E images from adjacent slices and real IHC images as references. Other cell types were also examined to identify incorrect and non-specific staining in cellular components and background. For these assessments, pathologists recorded their evaluations as "Yes" or "No". We calculated the percentages for each category based on the answers to demonstrate the consistency of the assessment across the pathologists. Additionally, pathologists were asked to rate their confidence (Very, Moderate, and Not Confident) in using the virtual images for NSCLC subtyping. In this regard, we calculated the overall percentage of pathologists' confidence in using virtual staining for diagnosis. Their diagnostic decisions were then compared to those made using real IHC images to evaluate consistency and reliability.

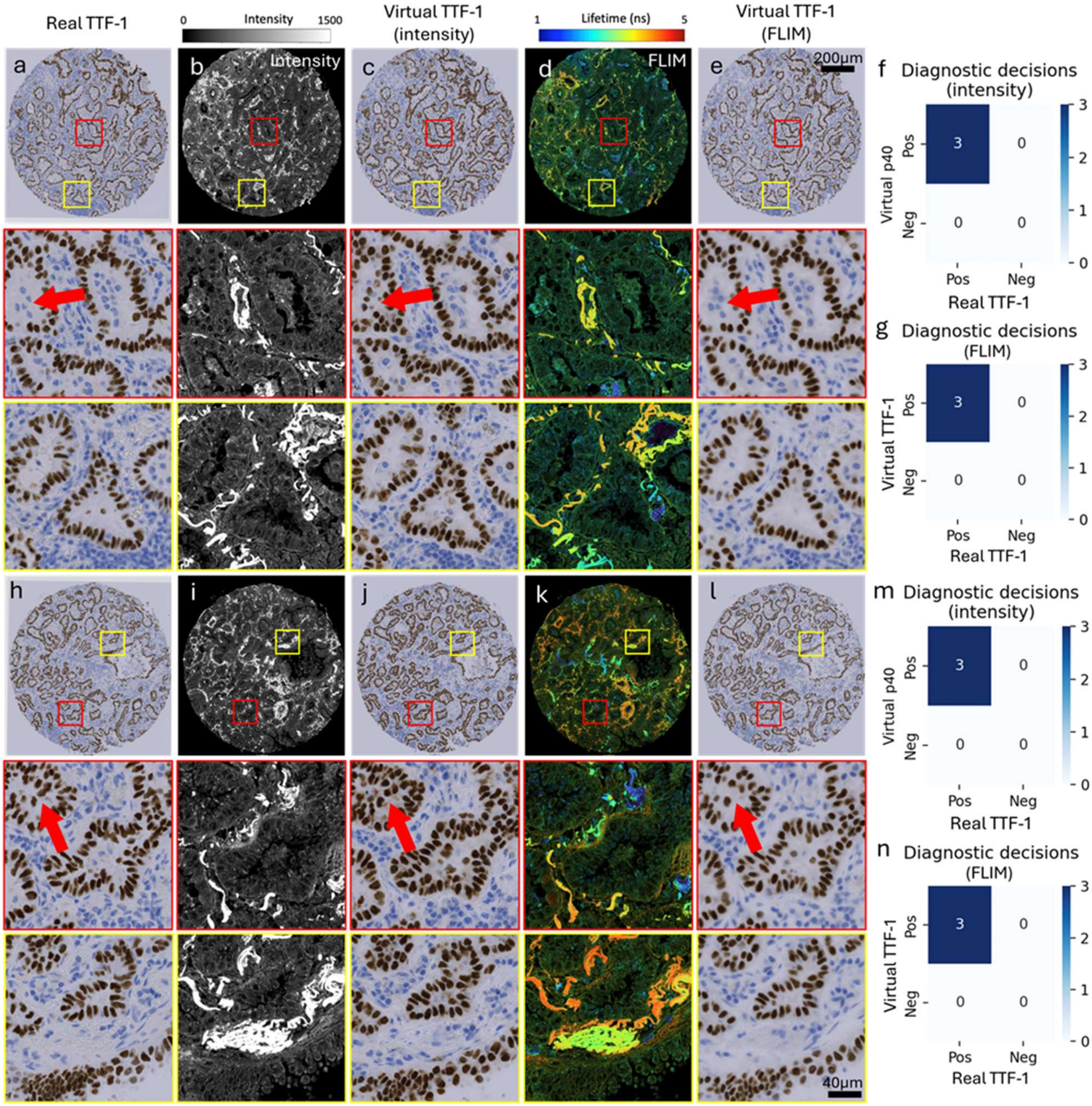

***Figure 4: Virtual TTF-1 staining on two TMA cores. a-g****: Virtual TTF-1 images from core 1, where both intensity-based (b) and FLIM-derived virtual images (c, e) closely resemble real TTF-1 staining (a), enabling consistent lung AC diagnosis by pathologists (f, m).* ***h-n****: Virtual TTF-1 images from core 2. Compared with real TTF-1 staining (h), both intensity- (i) and FLIM-derived (k) images produce virtual staining (j, l), which is suitable for lung AC diagnosis (m, n). However, the red arrows indicate some mis-reconstructed cells, where the intensity is inferior to FLIM in accurately reconstructing TTF-1+ cells.*

Fig. 4 presents virtual TTF-1 staining on two TMA cores, where both are TTF-1+ cases, indicating lung AC. Fig. 4a-g are for core 1 and Fig. 4h-n are for core 2, including both intensity and FLIM-based derivation. Within the figure, the presented FLIM images are the false-colour lifetime images with normalised intensity as the alpha channel[29]. In general, both modalities produce satisfactory outcomes (Fig. 4c, e, j, l) that closely resemble real TTF-1 images (Fig. 4a, h). All pathologists are confident in making accurate lung AC diagnoses using the virtual images (Fig. 4f, g, m, n), underscoring their reliability for robust clinical decision-making. However, discrepancies exist, as highlighted by red arrows in Fig. 4, which indicate instances where TTF-1+ cells are mis-reconstructed in intensity-based virtual images but accurately reconstructed in FLIM-based images. These underscore subtle differences in reconstruction accuracy between the two modalities, demonstrating that FLIM-based reconstruction generally outperforms intensity-based reconstruction. This becomes more obvious for TTF-1 cases. A special case (Supplementary Fig. 5) further demonstrates that intensity-based staining lacks clarity, making it challenging for pathologists to make confident diagnostic decisions. In contrast, FLIM-based reconstruction does not introduce the ambiguity.

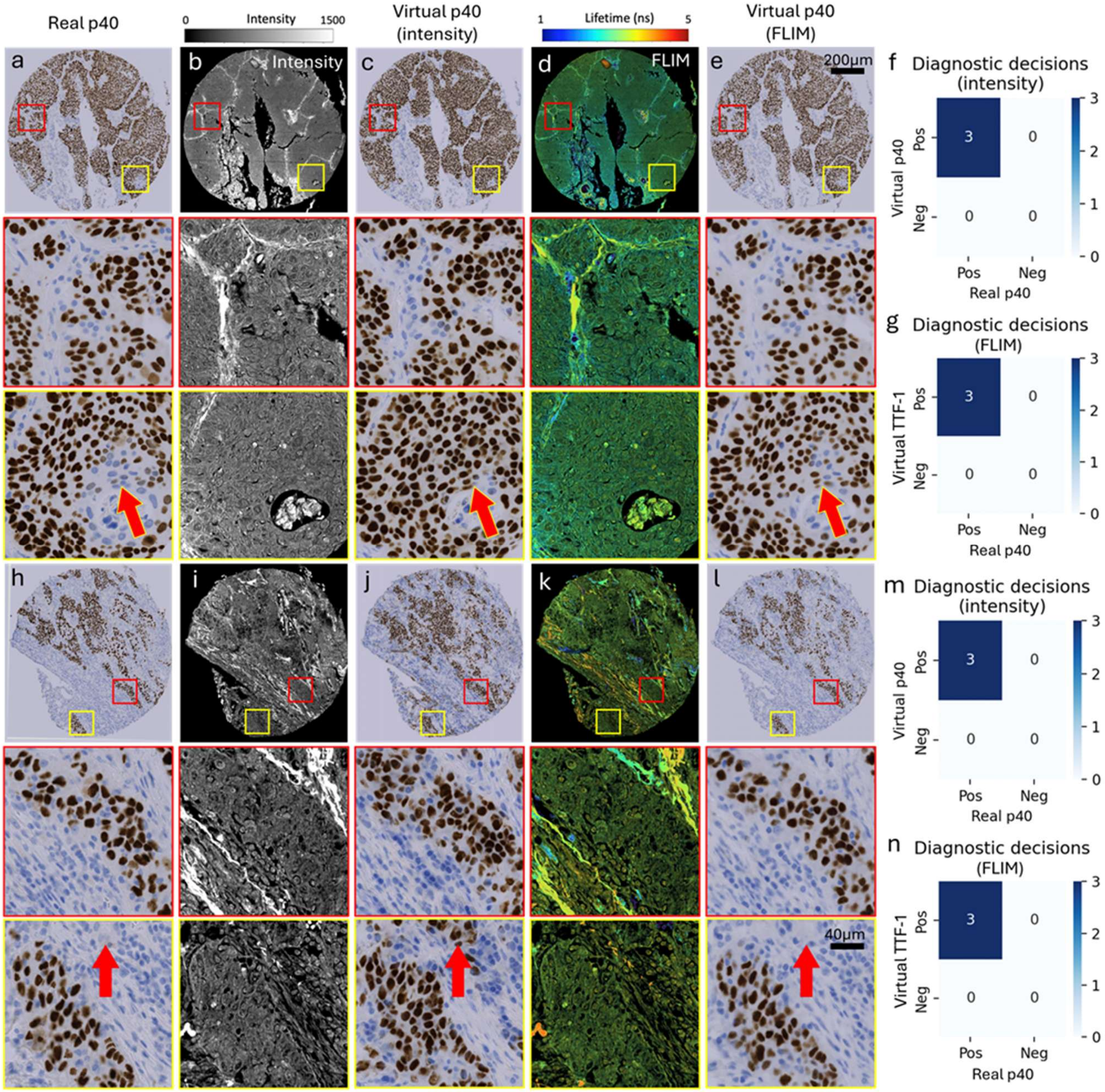

***Figure 5: Virtual p40 staining on two TMA cores. a-g:*** *Virtual p40 images from core 1, where both intensity-based (b) and FLIM-derived (c, e) reconstructions exhibit high fidelity to real p40 staining (a), enabling pathologists to reliably diagnose lung SqCC (f, m).* ***h-n****: Virtual p40 images from core 2. Relative to real p40 staining (h), virtual staining (j, l) is achieved using both intensity-based (i) and FLIM-derived (k) images, providing a reliable basis for lung SqCC diagnosis (m, n). However, red arrows indicate regions of mis-reconstructed cells, where intensity-based imaging demonstrates inferior accuracy compared to FLIM in reconstructing p40+ cells.*

Fig. 5 illustrates the results of virtual p40 staining using both intensity and FLIM. Like virtual TTF-1 staining, both intensity and FLIM-based approaches can generate high-quality virtual p40 images (Fig. 5c, e, j, l) for consistent clinical decision-making by pathologists (Fig. 5f, g, m, n). Red arrows reveal some differences in reconstruction accuracy between intensity- and FLIM-derived virtual p40 images. Intensity-based reconstruction tends to overestimate the presence of p40+ cells, whereas FLIM-derived reconstruction more accurately captures the true distribution of p40+ cells. These differences highlight the superior precision of FLIM-based imaging in faithfully identifying p40+ cells. Supplementary Fig. 6 illustrates a unique case, the only one in this study, where both intensity- and FLIM-based virtual p40 images fail to provide sufficient clarity, preventing pathologists from making highly confident diagnoses. However, this is mainly due to the ambiguity in the real p40 image, where artefacts may be introduced during the staining process. Nevertheless, both approaches enable pathologists to conduct consistent assessments for lung SqCC diagnosis, ensuring reliable clinical evaluations despite differences in reconstruction accuracy.

In clinical practice, the initial assessment of H&E-stained images is used to determine if the morphological features are sufficient to reach a definitive diagnosis of AC or SqCC[59]. However, when distinct morphological features are absent, immunohistochemical staining is required. Supplementary Fig. 7 shows a poorly differentiated solid pattern NSCLC, where H&E–based morphological staining alone is inconclusive for subtyping. In such cases, IHC staining provides a definitive diagnosis, and virtual IHC staining offers a rapid, non-destructive alternative for accurate diagnosis. We demonstrate concordance of both p40 expression in tumour cells and TTF-1 expression in benign pneumocytes. This facilitates the diagnosis of squamous cell carcinoma and underscores the utility of this approach in revealing expected patterns of protein expression in both benign and malignant cell populations. Quantitative analysis through histograms demonstrates that those cells can be differentiated from each other by average intensity (Supplementary Fig. 7l) and lifetime (Supplementary Fig. 7m), highlighting the potential to utilise autofluorescence signals for cell differentiation. The intensity and lifetime value distributions of all the samples are visualised in Supplementary Fig. 8.

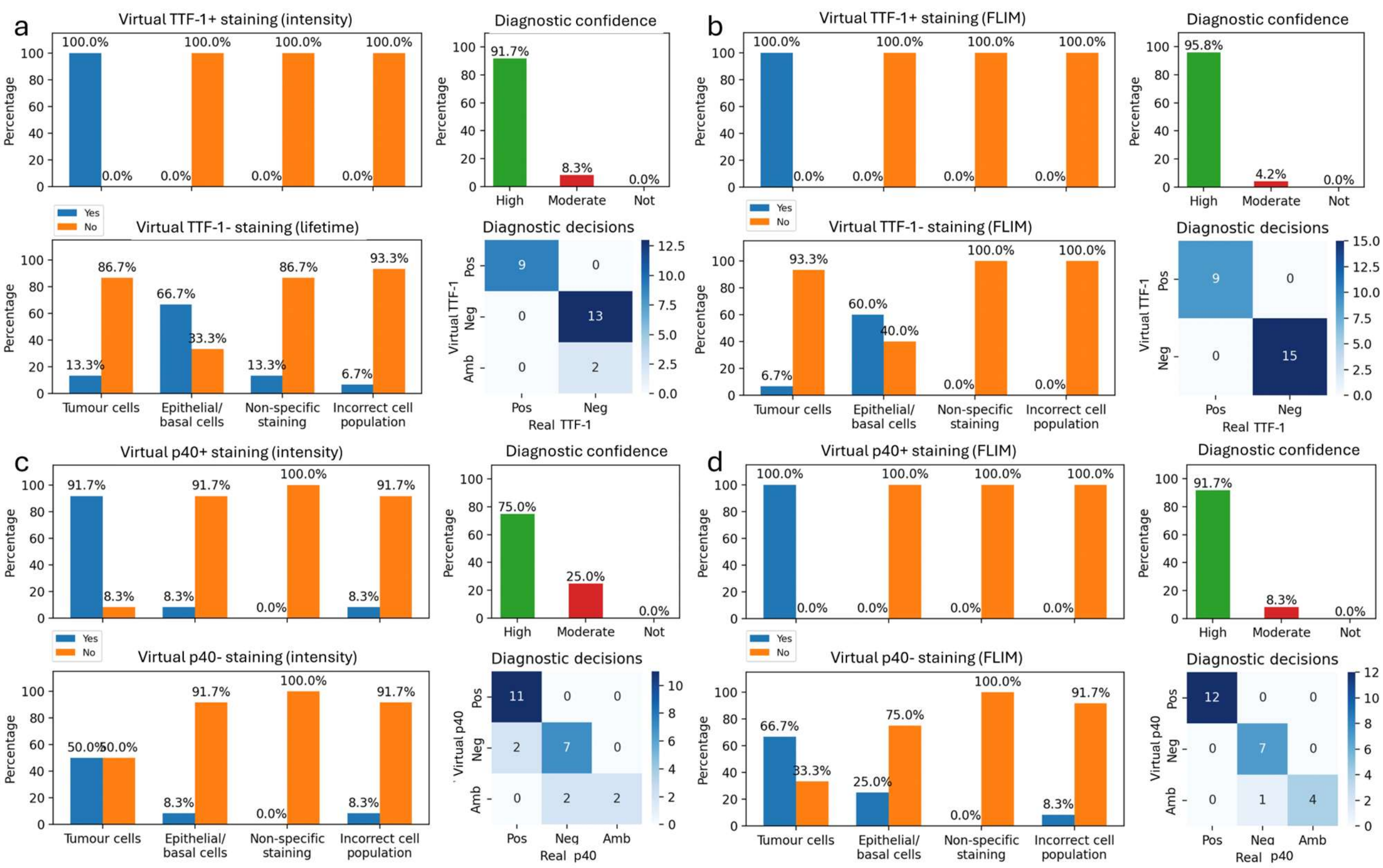

***Figure 6: Pathologist evaluation of virtual TTF-1 and p40 staining compared to corresponding true IHC images***. ***a***. *Intensity-based virtual TTF-1 staining.* ***b***. *FLIM-based virtual TTF-1 staining.* ***c***. *Intensity-based virtual p40 staining.* ***d***. *FLIM-based virtual p40 staining. For each case, pathologists assessed positive and negative expression in tumour cells, normal pulmonary epithelial/basal cells, and non-specific staining (including background and incorrect cell staining). Diagnostic confidence and decision-making were also evaluated to illustrate the overall quality of the methods for clinical decision-making.*

Fig. 6 presents the evaluation outcomes from pathologists, categorised by markers (TTF-1 and p40), imaging modalities (intensity and FLIM), and marker expressions (positive and negative). Fig. 6a and b show intensity- and FLIM-based virtual TTF-1 staining on eight cases (3 lung AC, 3 lung SqCC, and 2 OS), respectively. Overall, both intensity- and FLIM-based approaches produce satisfactory virtual TTF-1 images. For positive expressions, the reconstructed virtual images exhibit flawless staining details, with consistent scores across all metrics, enabling pathologists to make confident diagnostic decisions. In TTF-1-negative cases, although staining details were not perfectly reconstructed, particularly in intensity-based results, pathologists still performed accurate diagnoses, except in one negative case (Supplementary Fig. 5), where two out of three pathologists found it challenging to make confident decisions. The results in Fig. 6 clearly demonstrate the superiority of FLIM-based virtual TTF-1 staining over intensity-based methods, especially in negative cases, where consistent diagnoses were made on virtually stained images using FLIM.

Fig. 6c and 6d illustrate the inspection outcomes for virtual p40 staining in 8 NSCLC cases (4 SqCC, 3 AC, and 1 OS). Generally, both modalities can generate virtual p40 images suitable for clinical diagnosis. However, two cases exist where not all pathologists could make consistent decisions on real and virtual p40 images. This may be due to ambiguity in the real p40 staining (Supplementary Fig. 6). As with virtual TTF-1, all virtual p40+ images exhibit excellent diagnostic quality. However, 2 virtual p40 images are unclear, making confident decisions challenging. In both modalities, FLIM images are superior to intensity images in synthesis. Only one FLIM-based virtual image was considered ambiguous for decision-making, whereas the pathologists misinterpreted four intensity-based virtual p40 images.

*Table 1:* ***Quantitative comparison of virtual TTF-1 and p40 by intensity and lifetime images****. Results, presented as mean and standard deviation across all testing data, indicate that FLIM consistently outperforms intensity images in virtual IHC staining accuracy on each of these metrics, including mean-squared error (MSE), the normalised mutual information (MNI), the peak signal-to-noise ratio (PSNR), and the structural similarity index metric (SSIM). Lower MSE values correspond to better results, whereas higher NMI, PSNR, and SSIM (close to 1) values indicate better performance.*

| | | **MSE ↓** | **NMI ↑** | **PSNR ↑** | **SSIM ↑** |
|---|---|---|---|---|---|
| **TTF-1** | Lifetime | 0.11±0.03 | 1.15±0.01 | 22.98±2.31 | 0.72±0.06 |
| | Intensity | 0.14±0.04 | 1.12±0.01 | 20.45±2.15 | 0.63±0.06 |
| **p40** | Lifetime | 0.10/0.04 | 1.16/0.02 | 23.53/2.90 | 0.72/0.05 |
| | Intensity | 0.16/0.06 | 1.12/0.01 | 19.88/3.32 | 0.61/0.06 |

Table 1 presents quantitative comparisons of four widely used similarity metrics, including mean-squared error (MSE), normalised mutual information (NMI), peak signal-to-noise ratio (PSNR), and structural similarity index metric (SSIM). The results clearly indicate that lifetime surpasses intensity for virtual TTF-1 and p40 for all metrics, which is consistent with the visual evaluation of the virtual images.

Although our training for this cohort was based on TMA cores, we evaluated subtyping and virtual staining performance on real-world core needle biopsy specimens acquired during routine diagnostic work for suspected NSCLC. These biopsies are often heterogenous as they containing tumour and non-tumour regions being acquired for diagnostic purposes. Five needle core biopsy were included with detailed clinicopathological annotations presented in Supplementary Table 4. Entire biopsy sections were analysed, and multiple regions were extracted for subtyping and virtual staining evaluation. Representative specimens 1 and 5 are taken as examples due to the relatively high differentiation (Supplementary Table 4). Patch level results (Supplementary Fig. 10 and Fig. 11) demonstrate the model maintained high-confidence subtyping performance on biopsy specimens, suggesting diagnostic capability. However, the fidelity of the virtual IHC staining was reduced compared with conventional IHC, particularly in terms of both the proportion and intensity of positively stained cells. Although nuclei-level positive staining was observed, predicted nuclei were consistently larger than ground-truth IHC nuclei, indicating greater sensitivity to morphological variation. Similar behaviour was observed for SqCC biopsies, where accurate subtyping was achieved, but virtual p40 staining exhibited artifacts. This discrepancy is primarily due to domain-shift issues between TMA and biopsy tissue Subtyping performance was also lower in other biopsies, with varying prediction probabilities in specimens 2 ($N_{patches}$=257, 0.390±), 3 ($N_{patches}$=81, 0.041±), and 4 ($N_{patches}$=121, 0.286±). The preliminary independent test on biopsies using a TMA-core-trained DL model performs well on moderately different specimens but worse on poorly differentiated samples in biopsy specimens, indicating that robust biopsy-level performance will require inclusion of large-scale biopsy datasets during training.

## Discussion

This study demonstrated the feasibility of label-free lung cancer subtyping and evaluated it using autofluorescence intensity and lifetime images acquired from unstained NSCLC samples. Leveraging DL-based binary and multi-class classification, we can discriminate between non-cancerous lung tissue, AC, SqCC, and other NSCLC subtypes, with averaged AUC scores of 0.981 and 0.996 using label-free intensity and FLIM images, respectively. Furthermore, our virtual staining results demonstrated the ability to generate clinical-grade virtual p40 and TTF-1 images for lung cancer diagnosis, which is routinely used in clinical pathology practice. The NSCLC classifier and virtual staining can be used either independently or in combination. Both approaches enhance the efficiency of lung cancer diagnosis and support clinical decision-making.

Existing studies have shown that endogenous autofluorescence could be utilised for lung cancer subtyping based on statistical methods. However, it may not be effective for all cases due to the interpatient heterogeneity[61]. To assess the interpatient heterogeneity in our case, the intensity and lifetime value distributions of the samples are visualised in Supplementary Fig. 8. Overall, the intensity values of AC, SqCC, and OS exhibit high similarity, with closely aligned mean and standard deviation values, making subtype differentiation challenging using statistical metrics due to the high homogeneity. Regarding lifetime distributions, while normal and OS tissues demonstrate distinct separation from AC and SqCC, the latter two present highly overlapping distributions. Our proposed subtyping strategies, leveraging label-free intensity and lifetime images with DL, effectively mitigate the challenges posed by data homogeneity and achieve superior classification accuracy. Existing studies on cancer subtyping using H&E staining report average AUCs of approximately 0.97[5], 0.95[7], 0.80[8] to distinguish AC and SqCC, and 0.97[10] for AC, SqCC, and SCLC differentiation. The latest research[61] trained a large DL model on H&E-stained images to develop a universal cancer detection model for multiple cancer types. The model achieved an AUC of 0.979 for lung cancer and non-cancer detection. Another pathology foundation model[63] was proposed for detecting 19 cancer types from H&E-stained slides, with an AUC of 0.909 for lung cancer detection. Our label-free intensity-based model outperforms these methods, achieving a higher average AUC of 0.9859. Notably, unlike traditional approaches that rely on exogenous staining agents to enhance tissue morphology, our subtyping model utilises label-free endogenous autofluorescence images. Furthermore, our FLIM-based model further enhances accuracy, reaching an AUC of 0.9967. The superior performance of the FLIM-based model can be attributed to its ability to capture more functional details of the microenvironment over time. The greater variability in lifetime distribution than intensity proves advantageous for feature extraction during DL model training, thereby improving subtyping accuracy. Based on the analysis in Supplementary Fig. 2–4, FLIM imaging provides valuable insights into subtyping patches within a malignant core, whereas intensity and H&E-stained imaging primarily rely on morphological and histological features. This underscores the need for a quantitative investigation of lifetime changes in non-malignant tissue components within cancerous regions. A deeper understanding of these alterations could enable the detection of cancer with fewer tissue samples. Given that common AC subtypes, such as solid, lepidic, acinar, papillary, and micropapillary, present in our dataset and illustrated by H&E-stained examples with subtype distributions in Supplementary Table 5 and Supplementary Fig. 9, often exhibit morphological similarities, the proposed lifetime quantification approach may provide additional contrast to aid in their differentiation.

Apart from the models presented in this study, we also explored advanced DL architectures for computer vision, such as Transformer[63] and ConvNext[64], which incorporate advanced feature extraction backbones. Unfortunately, not all models converged without further fine-tuning. Therefore, it is worth exploring these advanced models with optimised parameter settings to assess their potential for improving accuracy. In binary and multi-class classification, the intensity-based model achieved lower accuracy than the other three methods in distinguishing between AC and SqCC. Adopting advanced DL architectures could be a future direction for distinguishing AC and SqCC. In time-domain FLIM systems, intensity imaging requires accumulating photon counts over a dwell time per pixel using a coarse-grained timer module. In contrast, fluorescence lifetime values are derived from fluorescence decay curves in the temporal dimension, which require a high-precision timer module to measure the time-of-flight of emitted photons and encode time-tagged photons into fluorescence decays. This process requires complex hardware and significant post-processing. The intensity-based model offers a balanced solution by addressing the trade-offs between the lengthy lifetime fitting process, the cost of complex FLIM systems, and the need for high-accuracy classification. This approach provides an effective alternative for scenarios where high-precision timer and sensor modules in FLIM systems are unavailable, ensuring robust classification performance without requiring the full complexity of traditional FLIM setups. In addition to confocal scanning intensity images, widefield scanning intensity images with multiple wavelength emissions could serve as an alternative input data source, significantly reducing data acquisition time. However, the wide-field scanned images exhibit coarse morphological features in intensity. Therefore, leveraging advanced DL models to enhance the performance of intensity-based DL models for lung cancer subtyping is a promising direction for future work.

Additionally, our GAN-based virtual staining is the first work to synthesise stained images for specific biomarkers for AC and SqCC, the most prevalent forms of NSCLC, using label-free autofluorescence images. Our label-free virtual IHC staining on two markers for lung cancer subtyping demonstrates the potential of virtual histological staining to go beyond the current state-of-the-art in autofluorescence-based virtual H&E and other common histological staining techniques[36]. In addition, the results also indicate that single-band autofluorescence images are more effective than multi-channel images used in the existing method[66]. Experienced pathologists' visual evaluations highlight the effectiveness of our methodology in converting autofluorescence images into virtual IHC images for diagnostic use. Since true IHC stains were generated in an accredited pathology laboratory, this indicates that our synthetic outcomes align with clinical standards.

Combined with virtual H&E staining[29], our techniques can now generate virtual H&E, TTF-1, and p40 images from a single autofluorescence image. By bypassing the traditional multi-step tissue processing procedures, our methods could provide these routine tests in minutes, without compromising the accuracy of clinical decision-making. The success of virtual IHC staining suggests that autofluorescence signals may vary across different tumour phenotypes, highlighting the efficiency of lung cancer subtyping using autofluorescence images. This is particularly effective when routine H&E staining alone is insufficient for differentiating NSCLC subtypes and additional IHC is required. Virtual IHC staining can therefore provide visual evidence of cellular protein expression and allow accurate tumour classification in a clinical context. While this study demonstrates strong diagnostic performance on TMA cores, further training sets will be required for core biopsies, as well as practical considerations such as imaging acquisition time, throughput, and integration into routine histopathology workflows, which were not evaluated, but will be critical to address in future studies to support clinical translation.

The virtual staining reused the DL technique described in our previous study[29]. This has several advantages. For example, the training does not require extensive hyperparameter tuning, making transfer learning straightforward without modification. This will also help simplify the integration of all these techniques into a unified platform that generates all these synthetic images in one run. We evaluated a more complex generative model, ResViT[66], which integrates Vision Transformer and residual convolutional blocks to enhance global texture learning. However, it underperformed in detecting p40- and TTF-1-positive cores, likely due to its larger model size and limited training data, leading to overfitting. While diffusion models have recently been explored as alternatives to GANs, their direct application to label-free virtual staining remains ineffective[29]. Recent research[67, 68] indicates that integrating advanced training strategies and architectural improvements can enhance the performance of diffusion models. Nevertheless, model optimisation is beyond the scope of this study, as GAN-based methods remain the most practical choice for label-free virtual staining due to their efficiency and lower computational cost.

In conclusion, our label-free NSCLC subtyping approach enables rapid, accurate diagnosis of NSCLC subtypes without the need for conventional tissue processing and staining.

# Methods

### *Ethical Approvals and TMA Construction*

The TMAs used in this study were approved by Lothian NRS Bioresource, Regional Ethics Committee (REC) numbers 15/ES/0094 and 20/ES/0061, with study references SR1208, SR1949, and SR2046. Application SR1208 was approved by the NHS Lothian Caldicott Guardian (reference CRD19031). The SR1208 TMA was constructed from consecutive patients undergoing curative resection surgery for NSCLC in a regional thoracic centre over two years. In this cohort, no patient received adjuvant immunotherapy in line with the standard of care at the time. An experienced pathologist annotated each resection block, and cores were taken and embedded into the TMA. For each patient case, one area of non-cancerous lung and three punches of tumour areas were taken and embedded into separate blocks. SR1949 TMA included selected cases to ensure a balance of adenocarcinoma (10 cases), squamous cell carcinoma (10 cases), other subtypes (large cell carcinoma, neuroendocrine, and carcinoid; 5 cases), and non-cancerous lung (5 cases). For cancer regions, TMA cores were taken in duplicate, and single punches from non-cancerous lung were embedded into separate blocks. SR2046 included an archival NSCLC cohort and included 86 lung cancer cases of varying subtype and mutational status, with triplicate TMA punches being taken from separate blocks. For each TMA block, slides were prepared by cutting 4-micron tissue sections on glass slides. For FLIM imaging, samples were deparaffinised and mounted with a coverslip. Following imaging, coverslips were removed with xylene incubation, and the same slides were transferred to NHS laboratories for subsequent staining.In an independent group, we identified core biopsies performed as part of the diagnostic pathway for patients (3 AC and 2 SqCC). For each case, three consecutive slides were cut from the blocks with FLIM imaging being performed, followed by TTF-1, p40, and H&E staining across the three slides.

### *TTF-1 and p40 Staining*

TMA sections were stained with antibodies to TTF-1 (Agilent; Clone: 8G7G3/1; Dilution 1:200) or p40 (Biocare Medical; Clone: BC28; Dilution 1:100) using IHC protocol F on the Leica Bond III Platform. Digital whole-slide images were captured using the Leica GT450 scanner at 40 × magnification. Bright-field TMA projects are imported into QuPath, where each core is individually identified and exported as an uncompressed histology image for co-registration with the corresponding FLIM image.

### *Data Acquisition and Processing*

Our subtyping and virtual staining approaches are based on a large-scale dataset comprising samples from across multiple cohorts. Intensity and FLIM image acquisition share the same imaging export setup. Images were acquired using a Leica STELLARIS 8 FALCON FLIM microscope with a 20×/0.75 NA objective. The pixel size was configured to 0.3001 $\mu$m. The excitation and emission wavelengths were set at 445 nm and [460, 640] nm, determined by a wavelength-by-wavelength scan of the tissue. After scanning, fluorescence lifetime images were reconstructed from the raw data using the multi-exponential fitting algorithm in Leica LAX-X software. The number of lifetime components in the fluorescence lifetime decay is determined by $\chi^2$, with four lifetime components being adopted for fitting as this configuration achieves the smallest $\chi^2$ for all cores. While exporting images, each core on a TMA was segmented into 512 × 512 pixels for each tile. Exported intensity images were filtered with a threshold of 10 photon counts to remove some background. Photon count range [0, 2000] and lifetime range [0, 5] ns were applied when exporting the images to achieve consistent visualisation. After exporting image tiles of intensity and FLIM images from the Leica LAX-X system, an ImageJ-embedded stitching method[69] was used to assemble the tiles into a complete image. Since some cancer tissues exhibit low photon emission, leading to dim intensity images, we enhanced the brightness of the normalised intensity images using histogram stretching within a constant range to improve feature extraction, following a strategy similar to our previously reported study[29].

Unlike subtyping, virtual staining involves an additional image co-registration process that aligns morphological structures across intensity and histology images. Due to differences in imaging modalities, pixel sizes between FLIM and brightfield images were standardised using bicubic interpolation in MATLAB. Additionally, an affine geometric transformation in MATLAB was applied to correct geometric distortions. The intensity enhancement strategy aligns with that used in subtyping, while the detailed co-registration process is documented in our previous study[29]. Co-registered FLIM images and real stained TTF-1 and p40 images were cropped into 256 × 256 patches and fed into the pix2pix model[70].

### *Deep Neural Network Details*

The subtyping and virtual staining neural networks are depicted in Supplementary Fig. 10. The subtyping DL model was trained on the EPSRC Tier-2 National HPC Service, Cirrus, hosted by EPCC. The Cirrus GPU cluster comprises 38 GPU nodes, each equipped with four NVIDIA Tesla V100-SXM2-16GB (Volta) GPUs. We utilised four nodes to train our models. For training, cross-entropy was used as the loss function, and SGD was employed as the optimiser, with a momentum coefficient of 0.9 and a weight decay coefficient of $10^{-4}$. The initial learning rate was set to 0.1, with a step decay of 20 epochs using a gamma coefficient of 0.1 to enhance convergence. Training was conducted for 100 epochs with a batch size of 330, requiring approximately three days per model. Distributed data parallelism was implemented to accelerate training. TIMM[71] was used to implement Resnet-50[56], Efficientnet-b0[57], and DenseNet-169[58]. We assigned the number of input channels to one for intensity images and four for stacked FLIM images. The number of output nodes in the final fully connected layer was two for binary classification and four for multi-class classification. Pre-trained models were retrieved from the Cirrus cluster and tested on a local NVIDIA RTX A5000 GPU. Performance evaluations of the three models are presented in Supplementary Table 6, with DenseNet-169 being the primary choice due to its superior performance.

For virtual IHC staining, we applied the method used in our previous virtual H&E staining study[29]. Specifically, we integrated the pix2pix[70] GAN with additional loss functions, including the Structural Similarity Index Measurement (SSIM)[72] and Style Loss[73]. The training was conducted on the EPSRC Tier-2 National HPC Service, Cirrus, hosted by EPCC. For TTF-1, 49 cores were used in the study, with 40 allocated to training and 9 reserved for testing. For p40, 50 cores were used in the study, with 40 cores allocated to training and 10 cores to testing. The models were trained for 50 epochs using transfer learning to accelerate convergence, with a batch size of 16. The initial weight decay was set to $10^{-4}$ and reduced by a factor of 10 every 15 epochs for both the generator and discriminator. Further details about the model and training process can be found in our previous study on virtual H&E staining[29].

### *Blind assessment of images*

Virtual IHC images were blind evaluated by three thoracic pathologists with over 30 years of combined experience, where intensity and lifetime-derived images were anonymised and mixed. The evaluation was conducted in 5 aspects, including:

- Total slide staining assessment:
    - Staining of tumour cells: Yes/No.
    - Staining of normal pulmonary epithelial/basal cells: Yes/No.
- Tumour staining assessment:
    - Intensity: Strong/Weak/Negative.
    - Proportion: Diffuse/Focal/Negative.
- Staining Quality assessment:
    - Background/non-specific staining: Yes/No.
    - Expression in incorrect cell populations (e.g., lymphocytes): Yes/No.
- Diagnostic confidence to use virtual IHC image compared to true IHC: Very/Moderate/Not confident.
- Diagnostic decision based on autofluorescence image alone:
    - The virtual IHC image: Positive/Negative/Ambiguous.
    - Is the decision the same as on the real IHC image: Yes/No.

Pathologists selected one option for all questions in the questionnaire based on the virtual IHC images and were blinded to each other's responses. The outcomes presented in the Section Results were statistically analysed according to the pathologists' selections.

## Data Availability

The authors declare that all data supporting the results of this study are available in the paper and the Supplementary Information section.

## Code Availability

The TIMM library implementing DL models for subtyping is available at https://github.com/huggingface/pytorch-image-models. The pix2pix model utilised for virtual IHC staining is available at https://github.com/phillipi/pix2pix. The Style Loss function is available at https://pytorch.org/tutorials/advanced/neural_style_tutorial.html. FLIM images were stitched using Fiji MIST stitching plugin (https://github.com/usnistgov/MIST). MATLAB® was used for affine transformation to co-register FLIM and true histology images.

## Acknowledgement

This study was partially funded by UoE Wellcome Institutional Translational Partnership Accelerator Fund and Cancer Research Horizons Seed Fund (PIII140), UoE Medical Research Council and Harmonised Impact Accelerator Accounts awards (MRC/IAA/015 and HIAA/037), Engineering and Physical Sciences Research Council (EPSRC) Grant Ref EP/S025987/1, NVIDIA Academic Hardware Grant Program, and ARA is currently supported by a UKRI Future Leaders Fellowship (MR/Y015460/1). The funders played no role in the study design, data collection, data analysis and interpretation, or this manuscript's writing. For open access, the author has applied a Creative Commons Attribution (CC BY) licence to any Author Accepted Manuscript version arising from this submission. The authors acknowledge the valuable comments from Dr Marta Vallejo at the School of Mathematics and Computer Science of Heriot-Watt University. We are grateful to the staff in the Department of Pathology, NHS Lothian and the Imaging Facility at the Institute of Regeneration and Repair, The University of Edinburgh (UoE).

## Author contributions

Q.W. conceived the research. Z.Z. and Q.W. collected and processed autofluorescence images. Z.Z. and Q.W. conducted experiments on deep classification, and Q.W. conducted experiments on virtual IHC staining. A.R.A., D.A.D., K.E.Q., and A.D.J.W. performed the clinical aspects of the study, including tissue collection and processing, IHC staining, and designing and conducting blind evaluations. J.R.H. provided expertise on signal processing and deep learning. Z.Z. and Q.W. prepared the manuscript, and all authors contributed to and approved the manuscript. Q.W. and A.R.A. supervised the research.

## Competing interests

Q.W. has 2 patent applications (UK patent application numbers: GB2319396.4 and GB 2405104.7) on the methods presented in this manuscript. Q.W. is currently employed by Prothea Technologies. A.R.A. is a founder shareholder and consultant for Prothea Technologies.

***Figure 1. Binary classification performance evaluation for three groups of cancer types: Cancer vs. Non-Cancer, AC vs. (SqCC + OS), and SqCC vs. OS.*** *(a) Subtyping overview. (b), (d), and (f) show ROC curves with AUC scores and confusion matrices for the three binary classifications based on intensity images. (c), (e) and (g) present the same evaluation metrics based on FLIM images, (h) and (i) present ROCs and confusion matrices subtyping AC and SqCC from intensity- and FLIM-based models.*

***Figure 2.*** *(a–c) Representative AC, SqCC, and OS cores are shown through intensity, FLIM, and H&E-stained images, along with probability maps inferred from intensity-based and FLIM-based models. The colour bar range spans from 0.8 to 1.0, with patches having probabilities below 0.8 displayed in white. (d–g) patch-based t-SNE clustering results derived from the last fully connected layer of the DNN models trained with (d) intensity and (f) FLIM images are presented. Both the intensity- and FLIM-based models exhibit four distinct class clusters with clear boundaries. Misclassified outliers from the intensity-based model are indicated by white arrows in (d), corresponding to (e) and (g), which show the colour-coded t-SNE plots of each class cluster for the intensity- and FLIM-based models, respectively.*

***Figure 3. Evaluation of the DL model for four-class multi-classification using FLIMs and intensity image datasets.*** *(a)-(c) and (h)-(j) indicate ROCs, confusion matrices, and core-based prediction probabilities (indicated by violin plots with means μ and standard deviations σ) from the DL model trained with FLIM and intensity images, respectively. (d)-(g) and (k)-(n) are examples of intensity and FLIM images of cases normal, AC, SqCC, and OS, indicated with corresponding four-class prediction probabilities. (o) and (p), and (q) and (r), are the core-based confusion matrices and ROC curves from the intensity- and FLIM-based models, respectively.*

***Figure 4: Virtual TTF-1 staining on two TMA cores. a-g****: Virtual TTF-1 images from core 1, where both intensity-based (b) and FLIM-derived virtual images (c, e) closely resemble real TTF-1 staining (a), enabling consistent lung AC diagnosis by pathologists (f, m).* ***h-n****: Virtual TTF-1 images from core 2. Compared with real TTF-1 staining (h), both intensity- (i) and FLIM-derived (k) images produce virtual staining (j, l), which is suitable for lung AC diagnosis (m, n). However, the red arrows indicate some mis-reconstructed cells, where the intensity is inferior to FLIM in accurately reconstructing TTF-1+ cells.*

***Figure 5: Virtual p40 staining on two TMA cores. a-g:*** *Virtual p40 images from core 1, where both intensity-based (b) and FLIM-derived (c, e) reconstructions exhibit high fidelity to real p40 staining (a), enabling pathologists to reliably diagnose lung SqCC (f, m).* ***h-n****: Virtual p40 images from core 2. Relative to real p40 staining (h), virtual staining (j, l) is achieved using both intensity-based (i) and FLIM-derived (k) images, providing a reliable basis for lung SqCC diagnosis (m, n). However, red arrows indicate regions of mis-reconstructed cells, where intensity-based imaging demonstrates inferior accuracy compared to FLIM in reconstructing p40+ cells.*

***Figure 6: Pathologist evaluation of virtual TTF-1 and p40 staining compared to corresponding true IHC images****.* ***a****. Intensity-based virtual TTF-1 staining.* ***b****. FLIM-based virtual TTF-1 staining.* ***c****. Intensity-based virtual p40 staining.* ***d****. FLIM-based virtual p40 staining. For each case, pathologists assessed positive and negative expression in tumour cells, normal pulmonary epithelial/basal cells, and non-specific staining (including background and incorrect cell staining).*

*Diagnostic confidence and decision-making were also evaluated to illustrate the overall quality of the methods for clinical decision-making.*

*Table 2:* ***Quantitative comparison of virtual TTF-1 and p40 by intensity and lifetime images****. Results, presented as mean and standard deviation across all testing data, indicate that FLIM consistently outperforms intensity images in virtual IHC staining accuracy on each of these metrics, including mean-squared error (MSE), the normalised mutual information (MNI), the peak signal-to-noise ratio (PSNR), and the structural similarity index metric (SSIM). Lower MSE values correspond to better results, whereas higher NMI, PSNR, and SSIM (close to 1) values indicate better performance.*

# Supplementary information

## Supplementary figures

**Figure 1.** Binary classification for four cases using different DNN models.

**Figure 2.** An example core of the AC subtype with intensity, FLIM, and H&E-stained images.

**Figure 3.** An example core of the SqCC subtype with intensity, FLIM, and H&E-stained images.

**Figure 4.** An example core of the OS subtype with intensity, FLIM, and H&E-stained images.

**Figure 5.** A TMA core where the intensity-based virtual TTF-1 image is ambiguous for pathologists to make confident decisions.

**Figure 6.** A TMA core where both virtual p40 images are ambiguous for pathologists to make confident decisions.

**Figure 7.** Virtual IHC staining of a solid pattern NSCLC and surrounding lung parenchyma

**Figure 8.** Distributions of means and standard deviations for individual cores of four subtypes, within the test datasets.

**Figure 9.** H&E-stained images of AC's five subtypes, solid, lepidic, acinar, papillary, and micropapillary, involved in the datasets.

**Figure 10.** Subtyping and virtual staining results of biopsy 101.

**Figure 11.** Subtyping and virtual staining results of biopsy 106.

**Figure 12.** Overview of deep learning architectures for lung cancer subtyping and TTF-1 and p40 virtual staining.

## Supplementary tables

**Table 1.** Clinical details on a patient level for the tissue microarrays used in this study.

**Table 2**. Number of patches and cores of each subtype in training, validation, and test datasets.

**Table 3.** Performance evaluation of classical deep learning architectures for multiple cancer type classification, using different metrics.

**Table 4.** Morphological AC subtypes used in this study.

**Table 5.** Ground-truth clinicopathological annotations of lung biopsy specimens, including histological diagnosis and differentiation pattern.

**Table 6.** Extensive Performance Evaluation of DenseNet for binary classification.

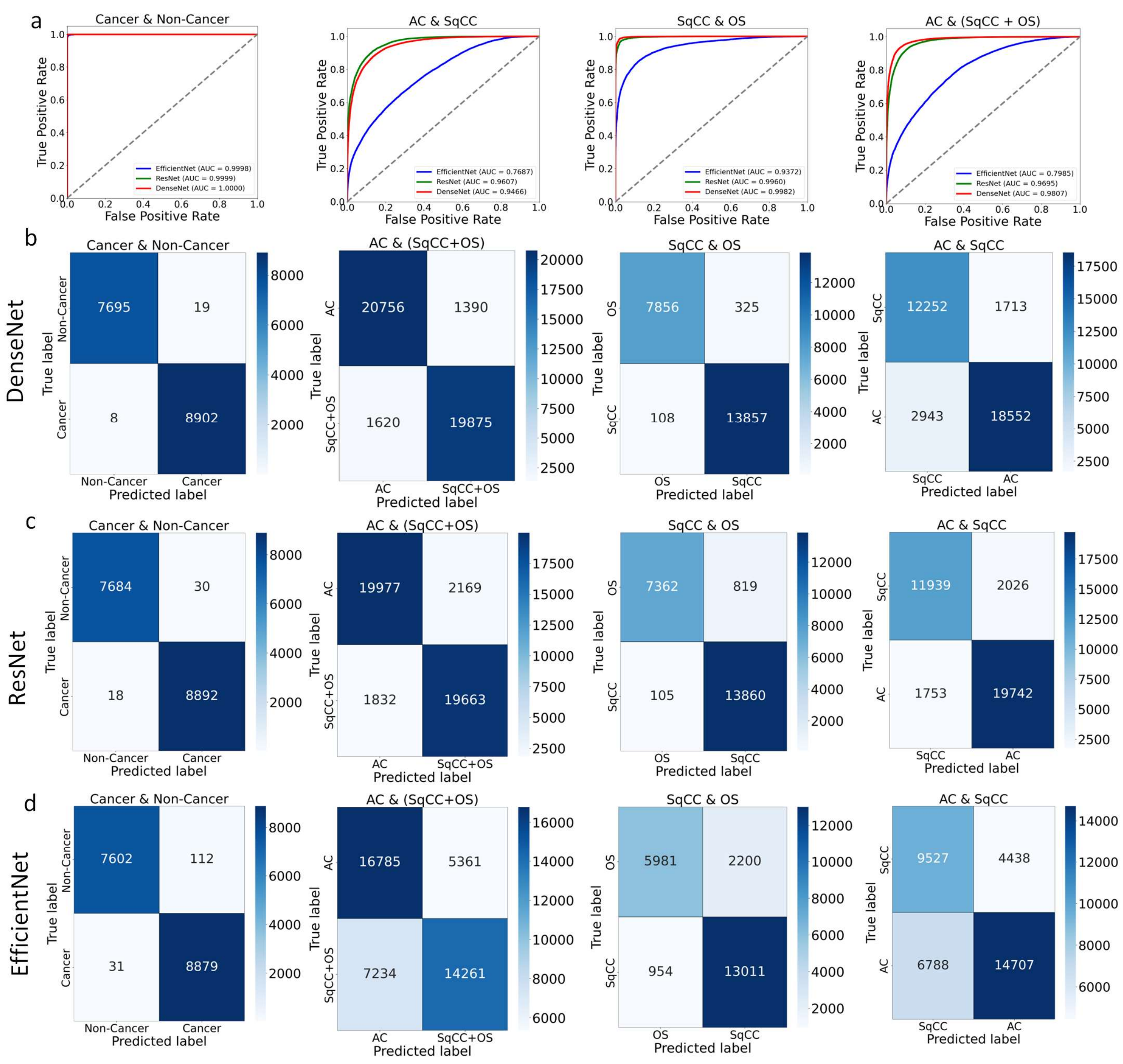

**Supplementary Fig. S1**. **Binary classification for four cases using different DNN models.** (a) ROCs and AUC scores, and (b), (c), and (d) confusion matrices from DenseNet, ResNet, and EfficientNet for four-class classification.

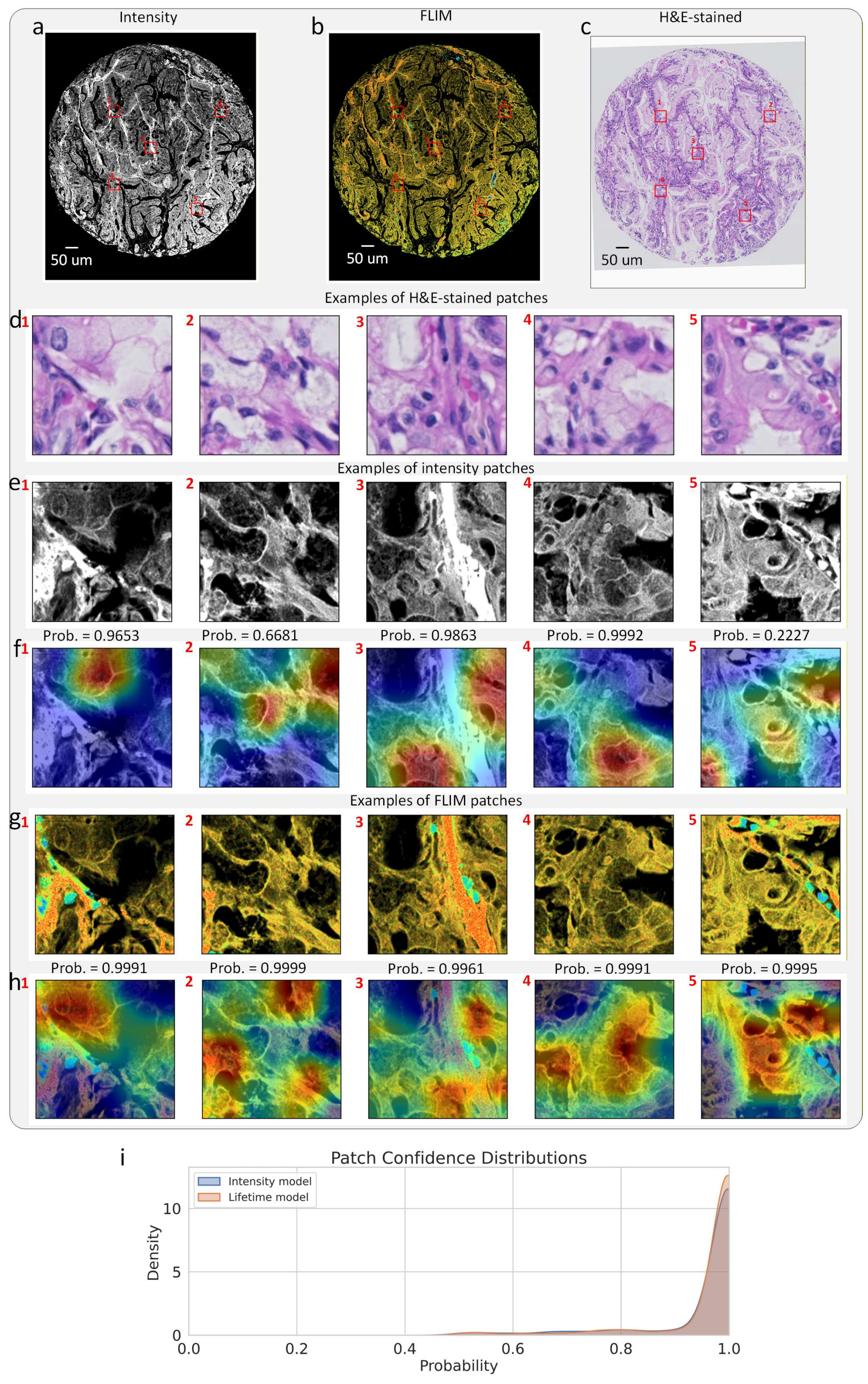

**Supplementary Fig. S2. An example core of the AC subtype with intensity, FLIM, and H&E-stained images**. (a-c) Five patches are highlighted as examples, showcasing different tissue components and morphologies. (d). H&E-stained images. (e) and (g) Inferred AC probabilities from the intensity- and FLIM-based DL model. (f) and (h) Salience maps generated by Grad-Cam++ from intensity- and FLIM-based models. (i) The probability distribution of intensity and lifetime models from patches in the core is presented.

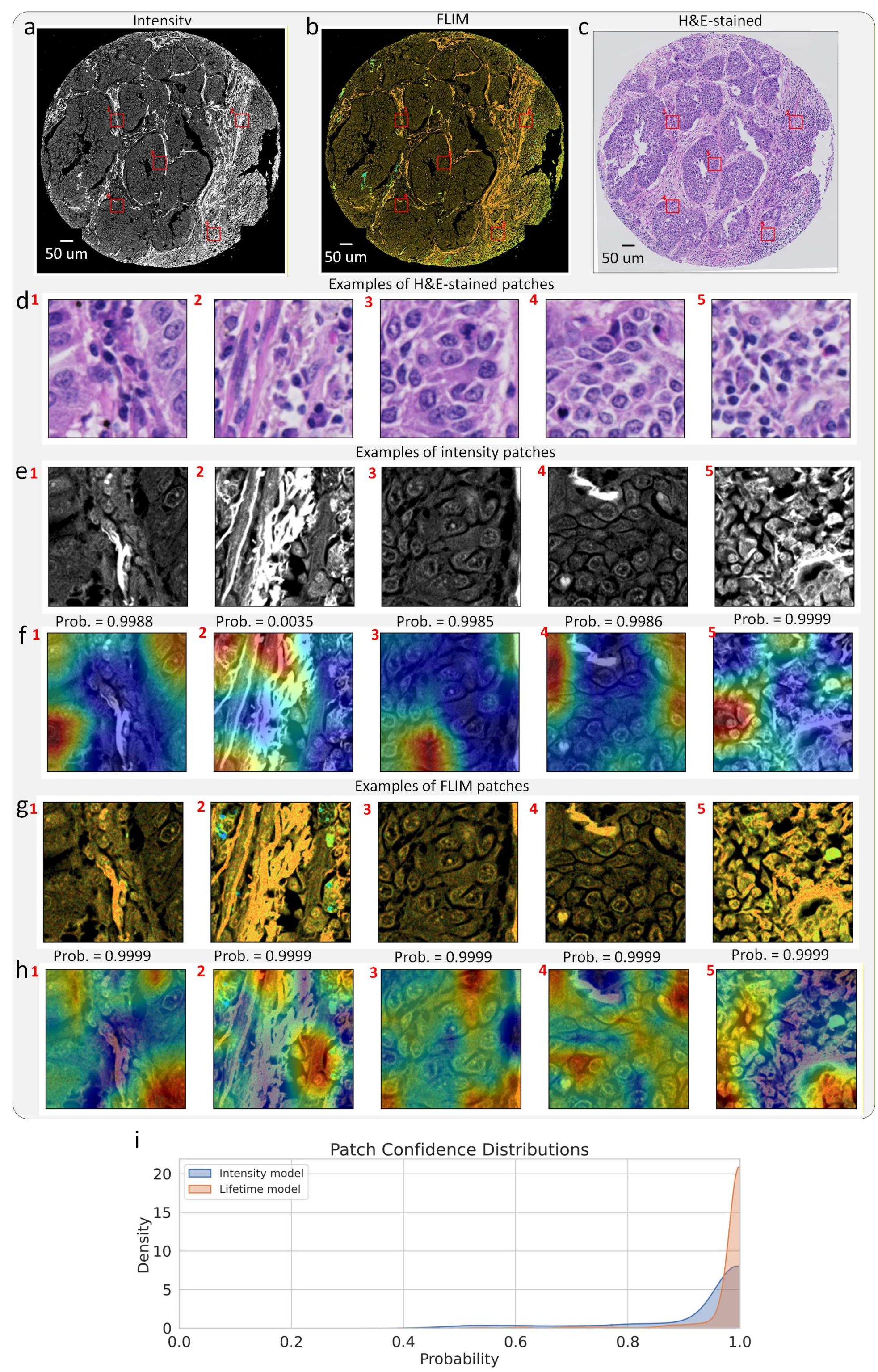

**Supplementary Fig. S3. An example core of the SqCC subtype with intensity, FLIM, and H&E-stained images.** (a-c) Five patches are highlighted as examples, showcasing different tissue components and morphologies. (d). H&E-stained images. (e) and (g) Inferred AC probabilities from the intensity- and FLIM-based DL model. (f) and (h) Salience maps generated by Grad-Cam++ from intensity- and FLIM-based models. (i) The probability distribution of intensity and lifetime models from patches in the core is presented.

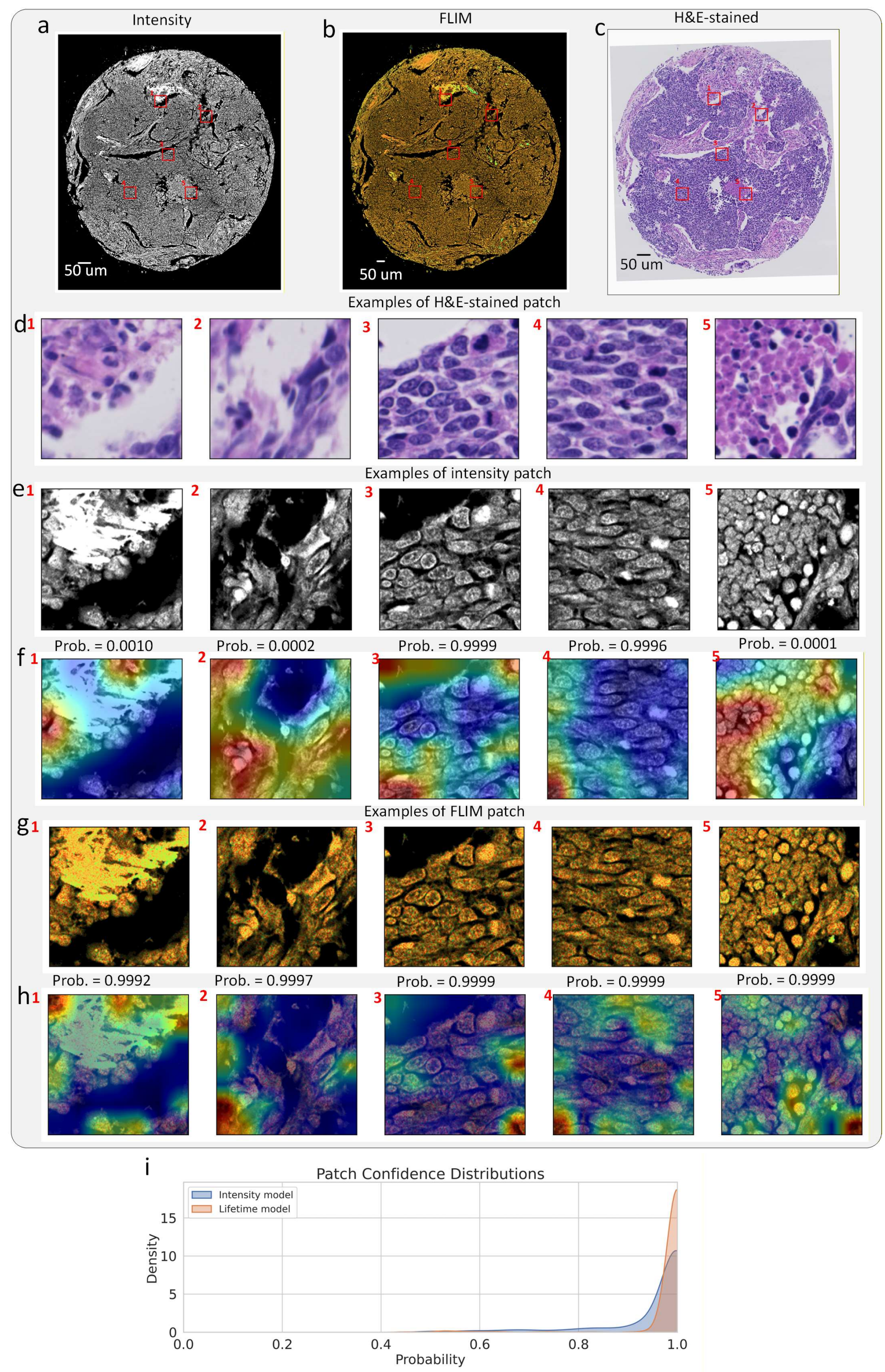

**Supplementary Fig. S4. An example core of the OS subtype with intensity, FLIM, and H&E-stained images.** (a-c) Five patches are highlighted as examples, showcasing different tissue components and morphologies. (d). H&E-stained images. (e) and (g) Inferred AC probabilities from the intensity- and FLIM-based DL model. (f) and (h) Salience maps generated by Grad-Cam++ from intensity- and FLIM-based models. (i) The probability distribution of intensity and lifetime models from patches in the core is presented.

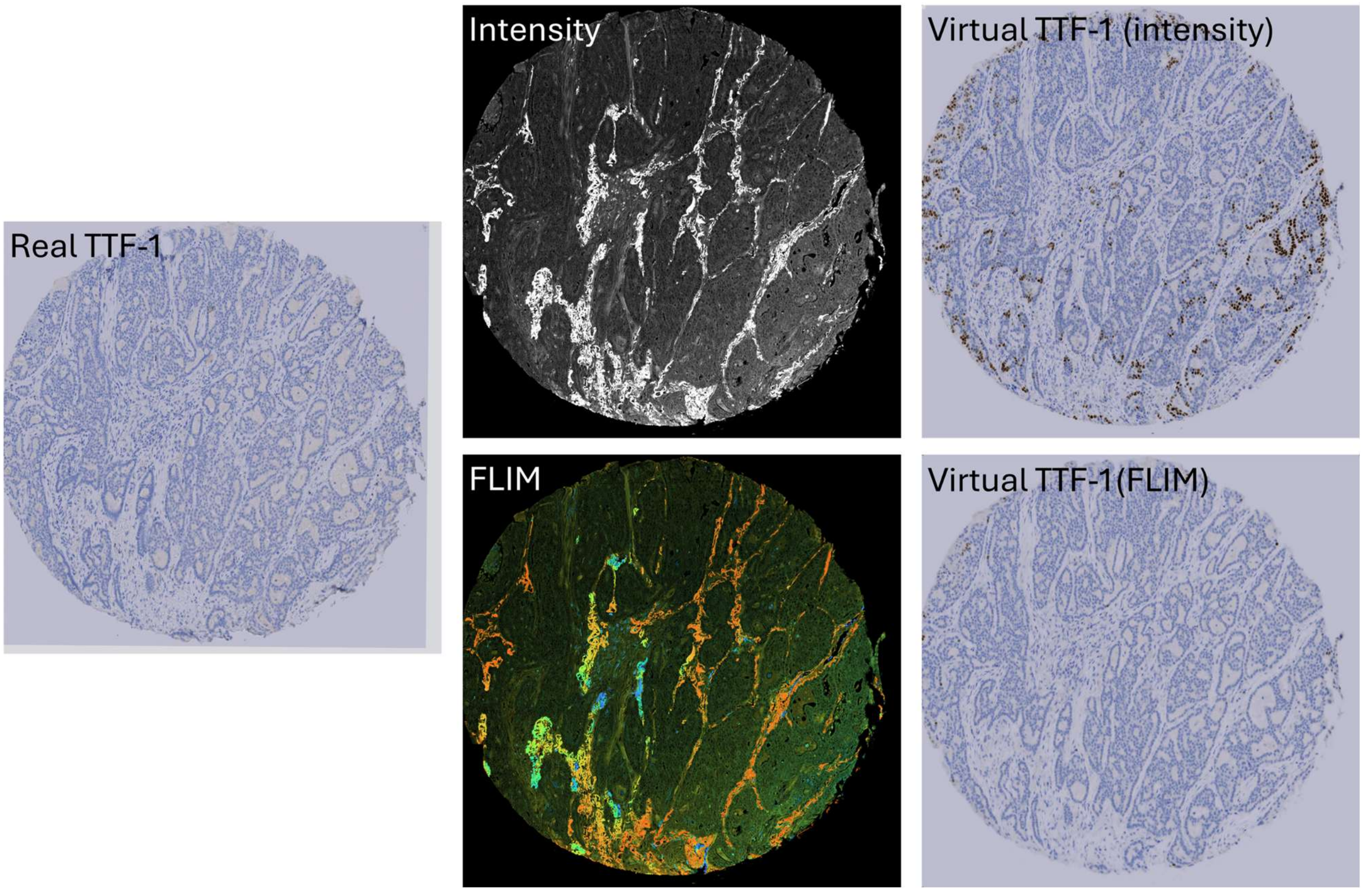

**Supplementary Fig. S5. A TMA core where the intensity-based virtual TTF-1 image is ambiguous for pathologists to make confident decisions.**

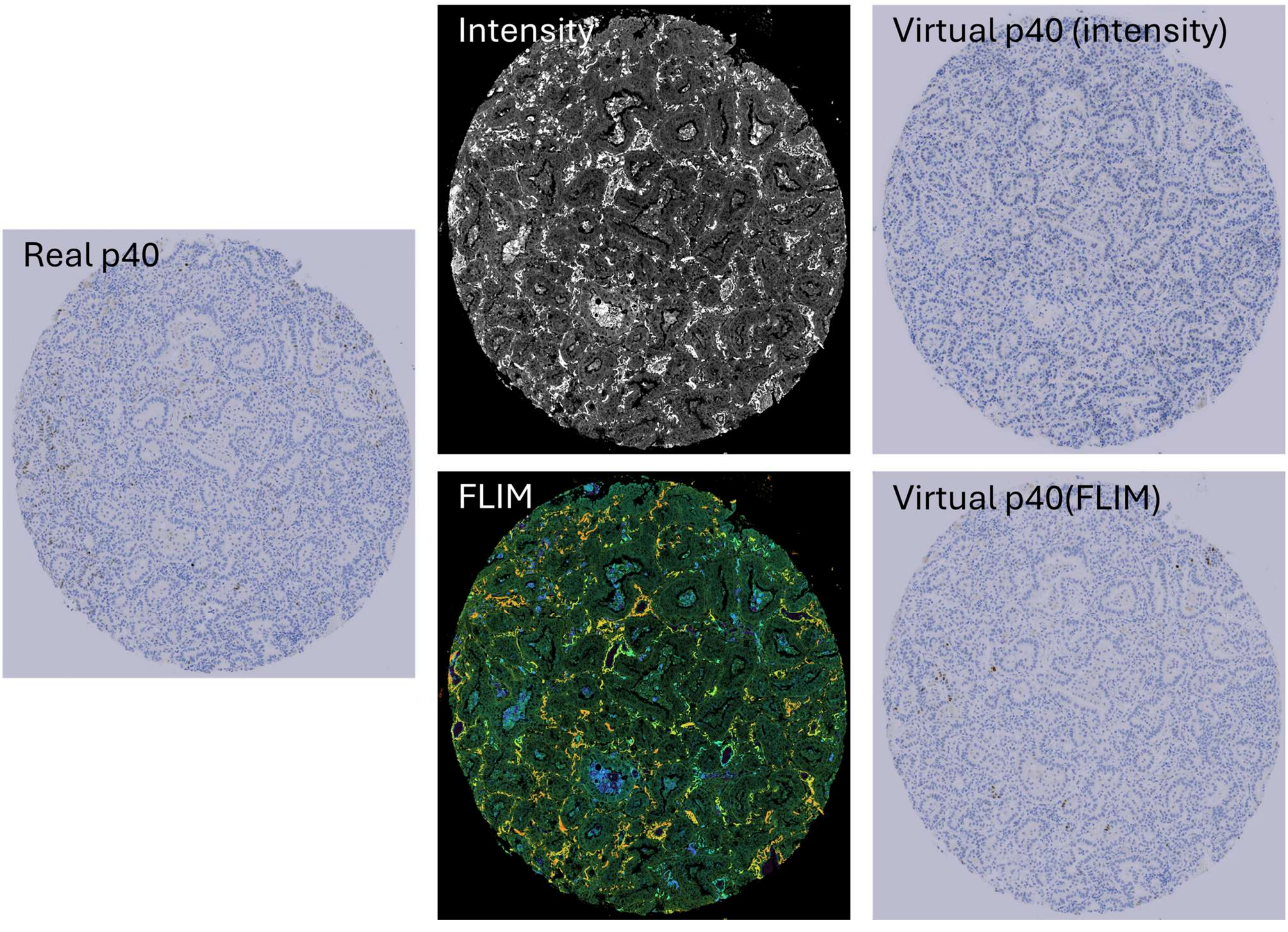

**Supplementary Fig. S6: A TMA core where both virtual p40 images are ambiguous for pathologists to make confident decisions.**

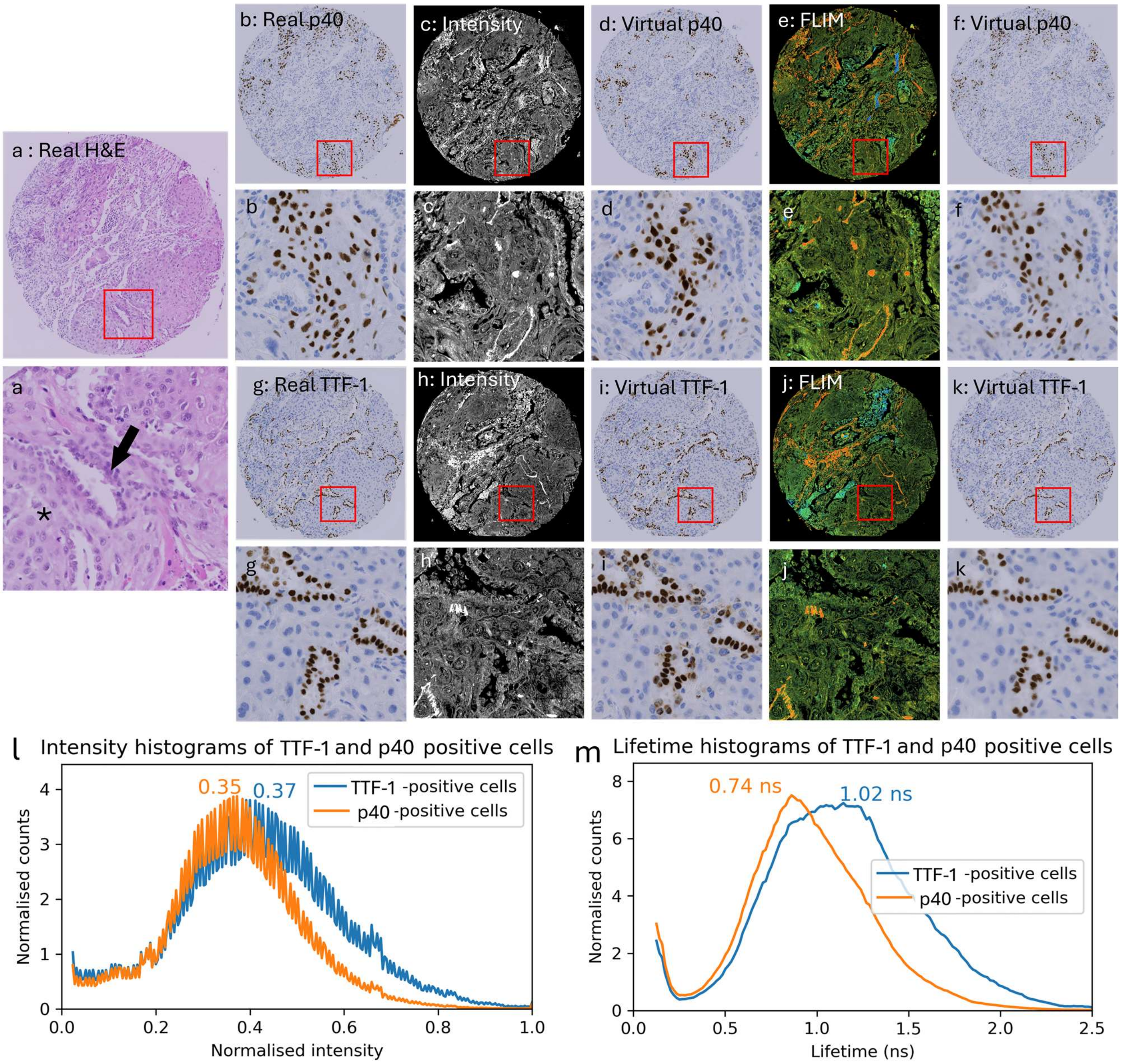

**Supplementary Fig. S7: Virtual IHC staining of a solid pattern NSCLC and surrounding lung parenchyma.** H&E stain (a) demonstrates solid nests of poorly differentiated malignant epithelial cells (*) with a squamoid morphology, which surround groups of entrapped, reactive alveolar pneumocytes (arrow). In a clinical context, p40 and TTF-1 IHC use is advisable for definitive subtyping. Here, both p40 (b) and TTF-1 (g) are expressed, with p40 expression within the malignant cells confirming the morphological impression of squamous cell carcinoma while TTF-1 expression highlights the benign, entrapped type 2 alveolar pneumocytes. Virtual p40 (d and f) and TTF-1 (i and k) images were synthesised from intensity (d from c and i from h) and FLIM (f from e and k from j) images, showing patterns of expression on an individual cell level almost identical to the ground truth (b vs d and f, and g vs i and k) in both the benign and malignant cell populations l and m show intensity and lifetime contrasts between TTF-1+ and p40+ cells, where histograms were generated from a cluster of those cells. The H&E, p40, and TTF-1 images were acquired from different sequential cuts of the same tissue block.

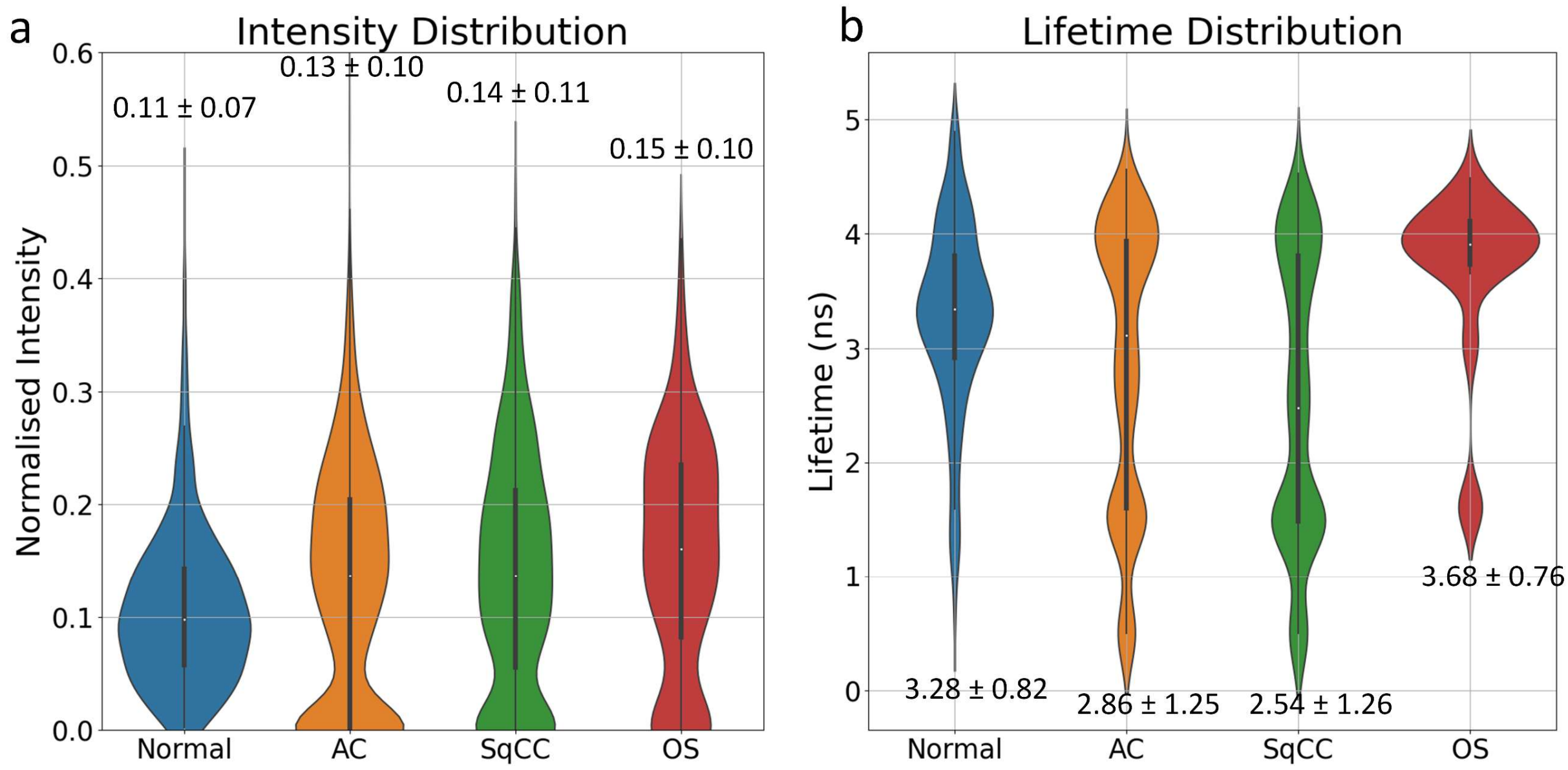

**Supplementary Fig. S8. Distributions of means and standard deviations for individual cores of four subtypes, within the test datasets.** (a) Normalised intensity value distributions. (b) Lifetime values distributions. The distribution of intensity values is more homogeneous than that of lifetime values.

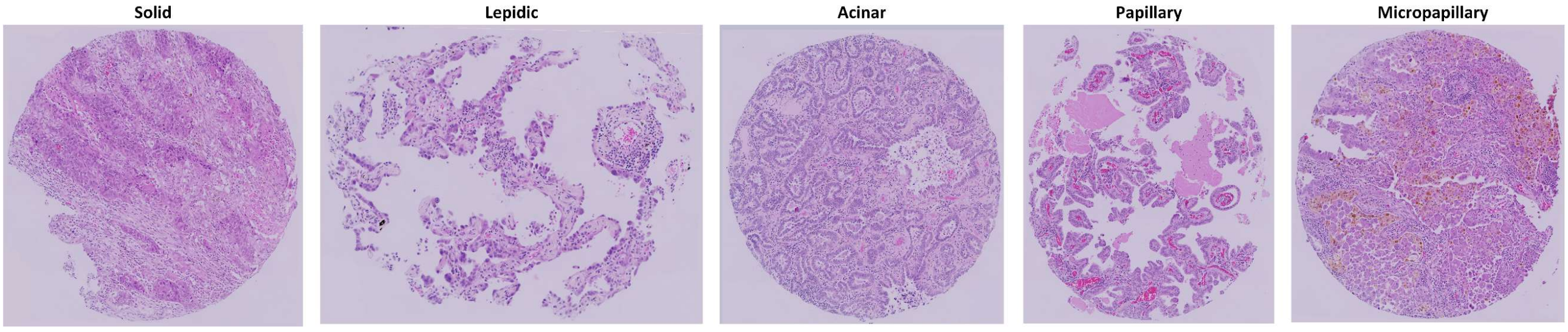

**Supplementary Fig. S9. H&E-stained images of AC's five subtypes, solid, lepidic, acinar, papillary, and micropapillary, involved in the datasets.**

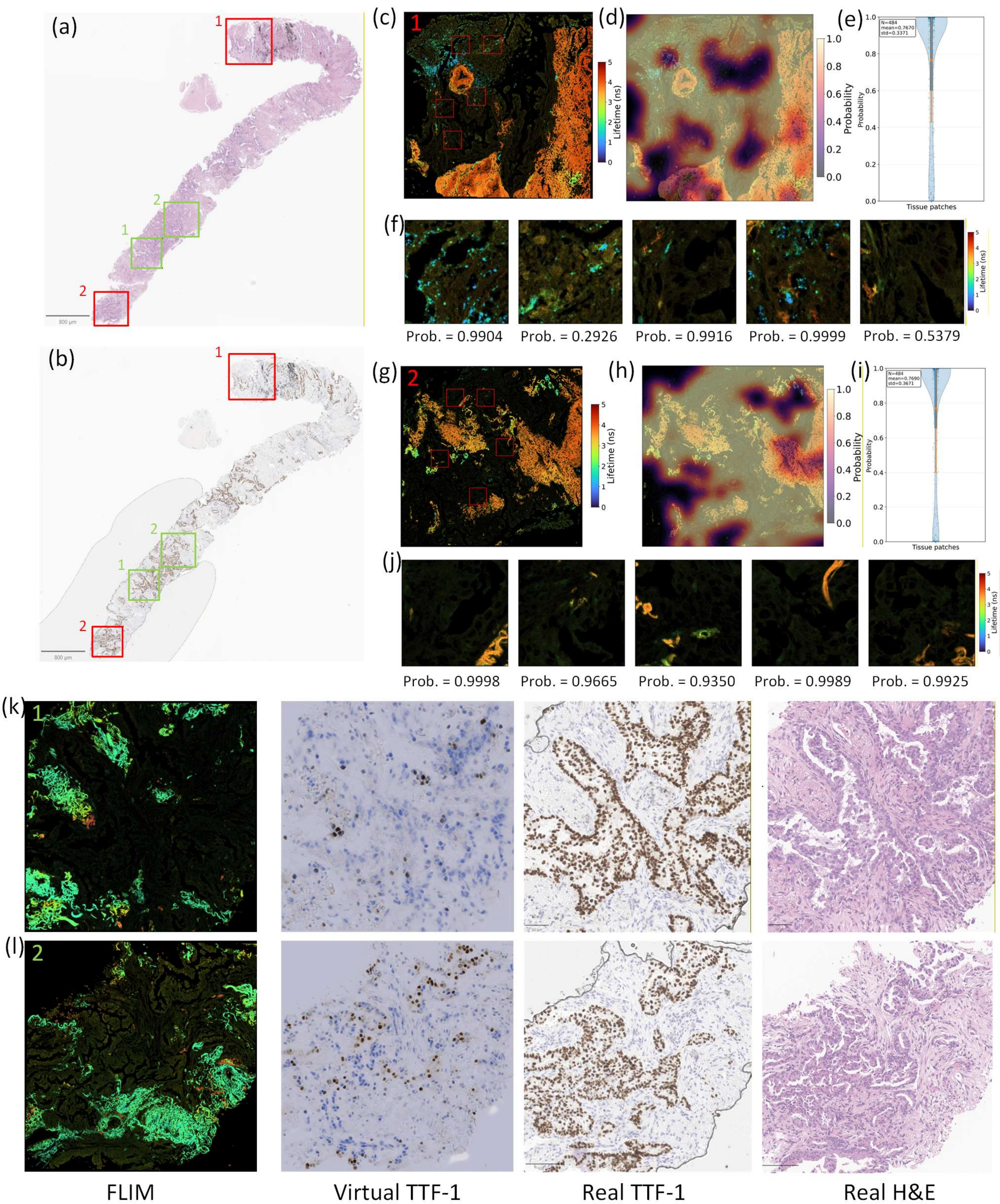

**Supplementary Fig. S10. Subtyping and virtual staining results of biopsy 101**. H&E overview of the ACC-annotated cancer specimen with regions of interest (ROIs) highlighted (red and green boxes for subtyping and virtual staining) and the corresponding IHC-stained slide are shown in (a, b). FLIM images of selected ROIs from regions 1 and 2 with patch-level sampling locations are shown in (c, g), together with DL predicted subtyping probability heatmaps overlaid on FLIM images in (d, h). The distributions of patch-level prediction probabilities within the ROIs are shown in (e, i), and representative FLIM patches with their corresponding predicted probabilities are shown in (f, j). Qualitative comparisons between virtual TTF-1 staining, real TTF-1 IHC, and real H&E histology for regions 1 and 2 are shown in (k, l).

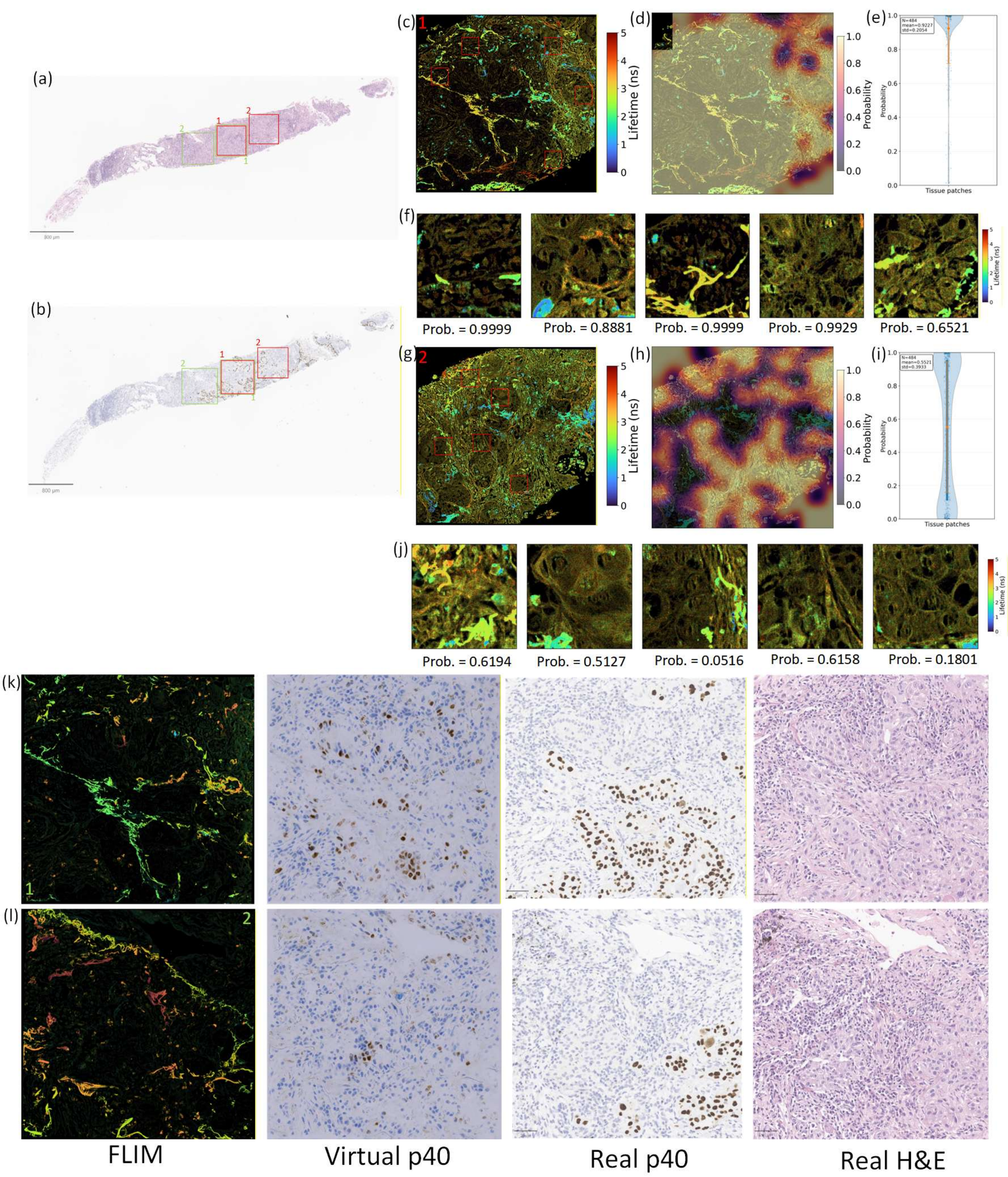

**Supplementary Fig. S11. Subtyping and virtual staining results of biopsy 106.** H&E overview of the SqCC-annotated cancer specimen with ROIs highlighted (red and green boxes for subtyping and virtual staining) and the corresponding IHC-stained slide are shown in (a, b). FLIM images of selected ROIs from regions 1 and 2 with patch-level sampling locations are shown in (c, g), together with DL–predicted subtyping probability heatmaps overlaid on FLIM images in (d, h). The distributions of patch-level prediction probabilities within the ROIs are shown in (e, i), and representative FLIM patches with their corresponding predicted probabilities are shown in (f, j). Qualitative comparisons between virtual p40 staining, real p40 IHC, and real H&E histology for regions 1 and 2 are shown in (k, l).

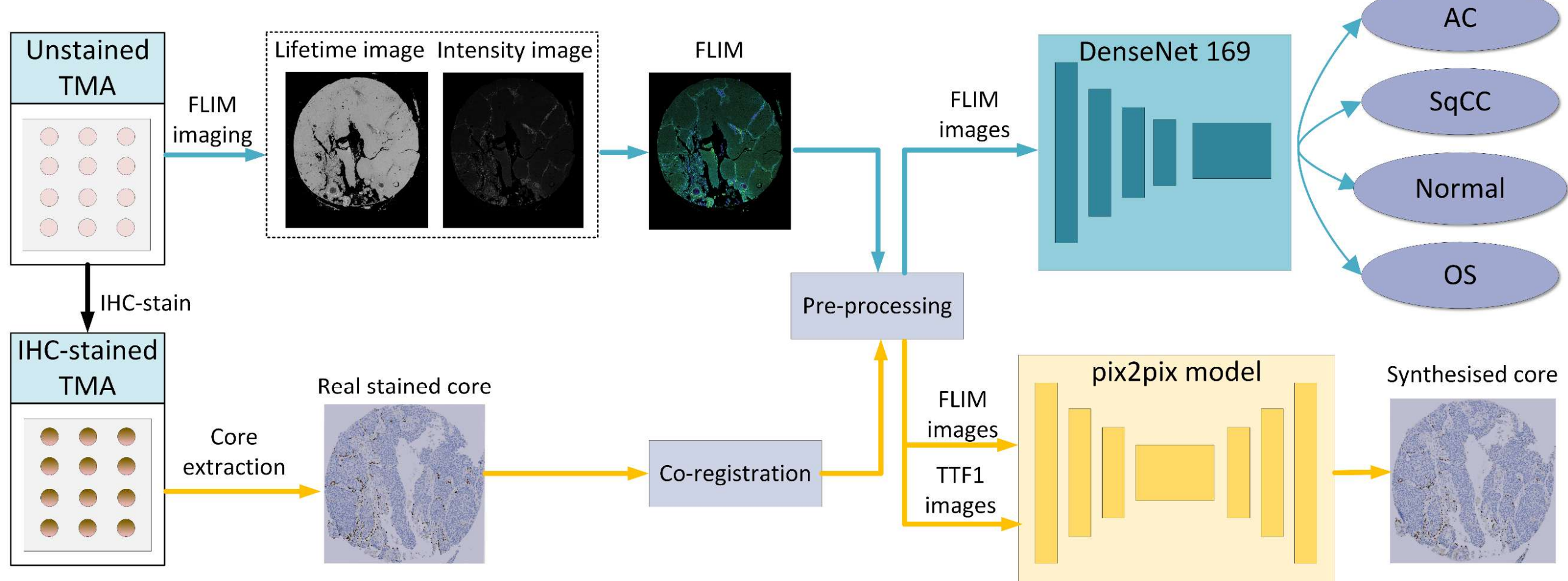

**Supplementary Fig. S12. Overview of deep learning architectures for lung cancer subtyping and TTF-1 and p40 virtual staining.** Blue and yellow arrows indicate the data processing pipelines for lung cancer subtyping and virtual staining, respectively. The subtyping pipeline consists of FLIM image generation, pre-processing to enhance contrast, and patch generation for downstream deep learning training and inference. The virtual staining pipeline includes co-registration to unify the morphology of FLIM and histology images. The pre-processing module generates paired FLIM and histology patches for training the network.

**Table 1. Clinical details on a patient level for the tissue microarrays used in this study.**

| **Demographics** | **TMA 1 (n=170)** | **TMA 2 (n=86)** | **TMA 3 (n=25)** |
|---|---|---|---|
| Male, n (%) | 72 (42.3%) | 40 (46.5%) | 12 (48%) |
| Female, n (%) | 98 (57.6%) | 46 (53.5%) | 13 (52%) |
| Age, mean (range) | 69 (46-92) | 67 (44-83) | 68 (52-83) |
| **Smoking Status** | | | |
| Current Smoker, n (%) | 77 (45.3%) | 13 (15.1%) | 8 (32%) |
| Ex-smoker, n (%) | 81 (47.6%) | 48 (55.8%) | 14 (56%) |
| Non-Smoker, n (%) | 10 (5.9%) | 25 (29.1%) | 3 (12%) |
| Unknown, n (%) | 2 (1.2%) | - | - |
| **Pathology** | | | |
| Adenocarcinoma, n (%) | 90 (52.9%) | 84 (97.7%) | 10 (40%) |
| Squamous Cell Carcinoma, n (%) | 61 (35.9%) | - | 10 (40%) |
| Other- Large Cell Carcinoma, n (%) | 7 (4.1%) | - | - |
| Other- Large Cell Neuroendocrine Carcinoma, n (%) | 4 (2.4%) | - | - |
| Other- Adenosquamous Carcinoma, n (%) | 6 (3.5%) | 2 (2.3%) | 1 (4%) |
| Other- Pleomorphic Lung Cancer, n (%) | 1 (0.6%) | - | - |
| Other- Adenocarcinoma with Large Cell Neuroendocrine, n (%) | 1 (0.6%) | - | 1 (4%) |
| Other- Squamous Cell Carcinoma with Large Cell Neuroendocrine, n (%) | - | - | 1 (4%) |
| Other- Carcinoid, n (%) | - | - | 2 (8%) |
| **Stage (TNM 7)** | | | |
| IA, n (%) | 54 (31.8%) | - | - |
| IB, n (%) | 54 (31.8%) | - | - |
| IIA, n (%) | 11 (6.5%) | - | - |
| IIB, n (%) | 23 (13.6%) | - | - |
| IIIA, n (%) | 21 (12.4%) | - | - |
| IIIB, n (%) | 7 (4.1%) | - | - |
| IV, n (%) | 0 (0%) | - | - |
| **Stage (TNM 8)** | | | |
| IA1, n (%) | - | 2 (2.3%) | 0 (0%) |
| IA2, n (%) | - | 7 (8.1%) | 5 (20%) |
| IA3, n (%) | - | 5 (5.8%) | 2 (8%) |
| IB, n (%) | - | 20 (23.3%) | 5 (20%) |
| IIA, n (%) | - | 7 (8.1%) | 0 (0%) |
| IIB, n (%) | - | 20 (23.3%) | 1 (4%) |
| IIIA, n (%) | - | 19 (22.1%) | 6 (24%) |
| IIIB, n (%) | - | 3 (3.4%) | 3 (12%) |
| IVA, n (%) | - | 2 (2.3%) | 1 (4%) |
| IVB, n (%) | - | 0 (0%) | 0 (0%) |
| N.A. | - | 1 (1.2%) | 2 (8%) |

**Table 2. Number of patches and cores of each subtype in training, validation, and test datasets**

| **No. patches** | | | | | |
|---|---|---|---|---|---|
| | **AC** | **SqCC** | **OS** | **Normal** | **Total** |
| **Training** | 96,147 | 82,094 | 42,820 | 41,428 | 262,489 |
| **Validation** | 21,328 | 16,656 | 7,148 | 7,904 | 53,036 |
| **Test** | 21,495 | 13,965 | 8,181 | 7,714 | 51,355 |
| **Total** | 138,970 | 112,715 | 58,149 | 57,046 | 366,880 |
| **No. cores** | | | | | |
| **Training** | 188 | 128 | 41 | 85 | 442 |
| **Validation** | 40 | 27 | 9 | 18 | 94 |
| **Test** | 40 | 27 | 10 | 18 | 95 |
| **Total** | 268 | 182 | 60 | 121 | 631 |

**Table 3. Performance Evaluation of classical deep learning architectures for multiple cancer type classification, using different metrics.**

| **Multi-Classification** | | | | | | |
| --- | --- | --- | --- | --- | --- | --- |
| | Accuracy Score | Precision Score | Specificity | Recall Score (Sensitivity) | Matthews Correlation Coefficient | AUC Score |
| **DenseNet-169** | **0.9619** | **0.9621** | **0.9855** | **0.9619** | **0.9458** | **0.9962** |
| EfficientNet-B0 | 0.8836 | 0.8852 | 0.9564 | 0.5046 | 0.8351 | 0.9925 |
| ResNet-50 | 0.9467 | 0.9469 | 0.9792 | 0.9467 | 0.9241 | 0.9864 |

**Table 4. Ground-truth clinicopathological annotations of lung biopsy specimens, including histological diagnosis and differentiation pattern.**

| Biopsy index | Diagnosis | Differentiation | Pattern | TTF-1 | p40 |
|---|---|---|---|---|---|
| 1 | ACC | Moderately | Acinar | Positive | Negative |
| 2 | ACC | Poorly | Solid | Positive | Negative |
| 3 | ACC | Poorly | Solid, focal acinar | Negative | Negative |
| 4 | SqCC | Poorly | Non keratinising | Negative | Positive |
| 5 | SqCC | Moderately | Keratinising | Negative | Positive |

**Table 5. Morphological AC subtypes samples used in this study.**

| No. Cores | | | | | | | | | | |
|---|---|---|---|---|---|---|---|---|---|---|
| **AC's subtypes** | **Solid** | **Solid and focal acinar** | **Acinar and Cribriform** | **Lepidic** | **Papillary** | **Acinar and Lepidic** | **Micropapillary** | **Acinar and Micropapillary** | **Acinar** | **Total** |
| **Amount** | 2 | 2 | 2 | 2 | 2 | 2 | 2 | 2 | 4 | 20 |

**Table 6. Extensive Performance Evaluation of DenseNet for binary classification.**

| | Accuracy | Precision | Specificity | Sensitivity | AUC score | MCC[1] |
|---|---|---|---|---|---|---|
| **Cancer & non-Cancer** | 0.9984 | 0.9979 | 0.9975 | 0.9991 | 0.9967 | 1.0000 |
| **AC & SqCC + OS** | 0.9310 | 0.9346 | 0.9372 | 0.9246 | 0.8621 | 0.9807 |
| **SqCC & OS** | 0.9804 | 0.9771 | 0.9603 | 0.9923 | 0.9580 | 0.9982 |
| **AC & SqCC** | 0.8935 | 0.9070 | 0.8550 | 0.9184 | 0.7762 | 0.9607 |

1. Matthews Correlation Coefficient